\documentclass[12pt]{spieman}  % 12pt font required by SPIE;
\usepackage{amsmath,amsfonts,amssymb}
\usepackage{graphicx}
\usepackage{setspace}
\usepackage{tocloft}
\usepackage{lineno}
\title{Overview and Reassessment of Noise Budget of Starshade Exoplanet Imaging}

\author[a,b,*]{Renyu Hu}
\author[a]{Doug Lisman}
\author[a]{Stuart Shaklan}
\author[a]{Stefan Martin}
\author[a]{Phil Willems}
\author[a]{Kendra Short}

\affil[a]{Jet Propulsion Laboratory, California Institute of Technology, Pasadena, CA 91109, USA}
\affil[b]{Division of Geological and Planetary Sciences, California Institute of Technology, Pasadena, CA 91125, USA}

\cftpagenumbersoff{figure}
\cftpagenumbersoff{table} 
\begin{document} 
\maketitle

\begin{abstract}
High-contrast imaging enabled by a starshade in formation flight with a space telescope can provide a near-term pathway to search for and characterize temperate and small planets of nearby stars.
NASA’s Starshade Technology Development Activity to TRL5 (S5) is rapidly maturing the required technologies to the point at which starshades could be integrated into potential future missions.
Here we reappraise the noise budget of starshade-enabled exoplanet imaging to incorporate the experimentally demonstrated optical performance of the starshade and its optical edge.
Our analyses of stray light sources -- including the leakage through micrometeoroid damage and the reflection of bright celestial bodies -- indicate that sunlight scattered by the optical edge (i.e., the solar glint) is by far the dominant stray light.
With telescope and observation parameters that approximately correspond to Starshade Rendezvous with {\it Roman} and HabEx, we find that the dominating noise source would be exozodiacal light for characterizing a temperate and Earth-sized planet around Sun-like and earlier stars and the solar glint for later-type stars.
Further reducing the brightness of solar glint by a factor of 10 with a coating would prevent it from becoming the dominant noise for both {\it Roman} and HabEx.
With an instrument contrast of $10^{-10}$, the residual starlight is not a dominant noise; and increasing the contrast level by a factor 10 would not lead to any appreciable change in the expected science performance.
If unbiased calibration of the background to the photon-noise limit can be achieved, Starshade Rendezvous with {\it Roman} could provide nearly photon-limited spectroscopy of temperate and Earth-sized planets of F, G, and K stars $<4$ parsecs away, and HabEx could extend this capability to many more stars $<8$ parsecs. Larger rocky planets around stars $<8$ parsecs would be within the reach of {\it Roman}.
To achieve these capabilities, the exozodiacal light may need to be calibrated to a precision better than 2\% and the solar glint better than 5\%. Our analysis shows that the expected temporal variability of the solar glint is unlikely to hinder the calibration, and the main challenge for background calibration likely comes from the unsmooth spatial distribution of exozodiacal dust in some stars.
Taken together, these results validate the optical noise budget and technology milestones adopted by S5 against key science objectives and inform the priorities of future technology developments and science and industry community partnerships.
%We provide an top-down overview of the system performance and noise budget of a starshade for exoplanet imaging and characterization. The required starlight suppression, straylight limitation, and other performance parameters are derived from aperture photometry and spectroscopy to detect and characterize Earth-sized exoplanets of nearby stars. We focus on the scalability of the noise budget per science objectives and the telescope that the starshade couples with. Specifically, we emphasize that xx.
\end{abstract}

% Include a list of up to six keywords after the abstract

% Include email contact information for corresponding author
{\noindent \footnotesize\textbf{*}Further author information: e-mail:\linkable{renyu.hu@jpl.nasa.gov} \\ @2020 California Institute of Technology. Government sponsorship acknowledged. }

\begin{spacing}{2}   % use double spacing for rest of manuscript

\section{Introduction}
\label{sec:intro}

Direct imaging of exoplanets from space holds promise to write a new chapter in astronomy and planetary science. With most of the exoplanets discovered to date in tightly bound orbits of their host stars, and thus uninhabitable unless the stars are much fainter than the Sun, direct imaging would detect planets in the habitable zones\cite{kasting1993habitable} of more Sun-like stars. If some of these planets are small and predominantly rocky in composition, they may have environments hospitable for life. One of the primary goals of exoplanet direct imaging is to search for temperate and small planets of nearby stars and study the chemical composition of their atmospheres with spectroscopy.

A starshade working in tandem with a space telescope provides one of the best near-term opportunities to achieve this goal. Starshade is an external occulter flown along the line of sight from a telescope to a target star. With the shape designed to mimic the optical effects of an optimally apodized screen, a starshade can create a ``deep shadow'' where the starlight is suppressed by $10^{10}$\cite{cash2006detection,vanderbei2007optimal}. The telescope kept in this shadow would be able to detect planets and disks around the star at very high contrast. A starshade that prevents the starlight from entering the telescope would allow many simplifications of the telescope optics. For example, precise wavefront control would not be necessary, which also reduces the number of reflections before feeding the light to a detector, and thus increases the optical throughput of the instrument. The costs of these benefits are the added complexity of formation flying and the complications involved in launching and deploying the large and optically precise starshade.

Two advanced mission concepts to discover Earth-like planets in the habitable zones of Sun-like stars being considered at NASA would use starshades as one of the starlight suppression techniques. The \textit{Roman Space Telescope}\cite{Spergel2015,Akeson2019} will be capable of collecting starlight reflected by large exoplanets with its coronagraph instrument. The \textit{Starshade Rendezvous Probe}\cite{Seager2019}, an advanced mission concept, would further enable {\it Roman} to search for Earth-sized planets in the habitable zones of $\sim10$ nearby stars, with the possibility to obtain their limited-bandwidth spectra at a moderate resolution ($\sim70$). The \textit{Habitable Exoplanet Observatory} (HabEx, \cite{Gaudi2020}), a concept of a 4-m space telescope with a starshade, has the main objective to search for Earth-sized planets in a larger stellar sample and obtain their spectra in a wider band and at a higher resolution. Starshade Rendezvous with {\it Roman} and HabEx, both with a starshade, would have the spectral characterization of small planets in the habitable zones of nearby stars as the key and probably limiting science objective.

To enable these potential exoplanet science missions, NASA’s Exoplanet Exploration Program (ExEP) is executing a directed and focused activity, the Starshade Technology Development Activity to TRL5 (S5). For S5, TRL5 is defined as demonstrating critical performance in relevant environments at the subsystem level with medium fidelity prototypes. The technology development plan of S5\footnote{https://exoplanets.nasa.gov/internal\_resources/1033/} adopts formation flying and observation scenarios of the Rendezvous and HabEx mission concepts as the baseline. Completion of S5 would bring starshade technologies to TRL5 for both concepts. Specifically, S5 includes experiments and analyses to demonstrate small-scale starshade masks that could reach $10^{-10}$ instrument contrast at the inner working angle (IWA) at a flight-like Fresnel number, to develop an optical edge for the starshade petals that would limit scattered sunlight (i.e., solar glint) to acceptable levels, to demonstrate the ability to sense the lateral offset between the starshade and the telescope to an accuracy of 30 cm, and to demonstrate the ability to design and manufacture the starshade mechanical elements that could meet the contrast requirement. Together with S5, ExEP has chartered a Science and Industry Partnership to engage the broader science and technology communities during the execution of the S5 activity.

Given the completion of most of S5's technology milestones on instrument contrast\cite{Harness2019a,Harness2019b}, solar glint\cite{Hilgemann2019}, and formation flying\cite{Flinois2018}, we are motivated to revisit the noise budget of starshades' application in exoplanet imaging. While estimates of exoplanet yields from starshade-assisted imaging have been published\cite{turnbull2012search,stark2016maximized,stark2019exoearth}, these works have not mapped the technology progress to science performance with the level of detail of the present paper. Particularly, previous works have not explicitly included solar glint in their noise budget. The purpose of this paper is to update the expected performance of starshade-enabled exoplanet imaging in light of new constraints on starlight and stray light suppression resulted from S5 work and to assess the noise budget of the spectral characterization of temperate and small planets of nearby stars. We will focus on the performance parameters that are {\it directly} related to exoplanet imaging, and defer the assessment on mechanical precision tolerance and stability -- which controls the instrument contrast -- to an error budget analysis\cite{shaklan2015error} and the S5 technology development plan. We will focus on revealing the dominating noise term under a wide range of realistic planet scenarios, to guide the priorities of future development. We will first evaluate the science performance on the assumption that the background could be calibrated to the photon-noise limit, and then discuss the impact of temporal and spatial variability on background calibration.

The paper is organized as follows. We first describe the model used to derive the S/N of a starshade-enabled exoplanet observation based on performance parameters in \S~\ref{sec:model}. In \S~\ref{sec:stray}, we discuss a range of stray light sources that may enter the telescope and are not included as a potentially dominant term in the noise budget. \S~\ref{sec:analysis} presents the expected S/N for observing nearby planetary systems with the current performance demonstrated by S5 and with potential future development. We discuss imperfect background calibration due to temporal and spatial variability and the sensitivity of exozodi levels in \S~\ref{sec:discussion}, and conclude with future prospects in \S~\ref{sec:conclusion}.

\section{STARSHADE PERFORMANCE MODEL}
\label{sec:model}

\begin{figure}[!h]
\centering
\includegraphics[width=1.0\textwidth]{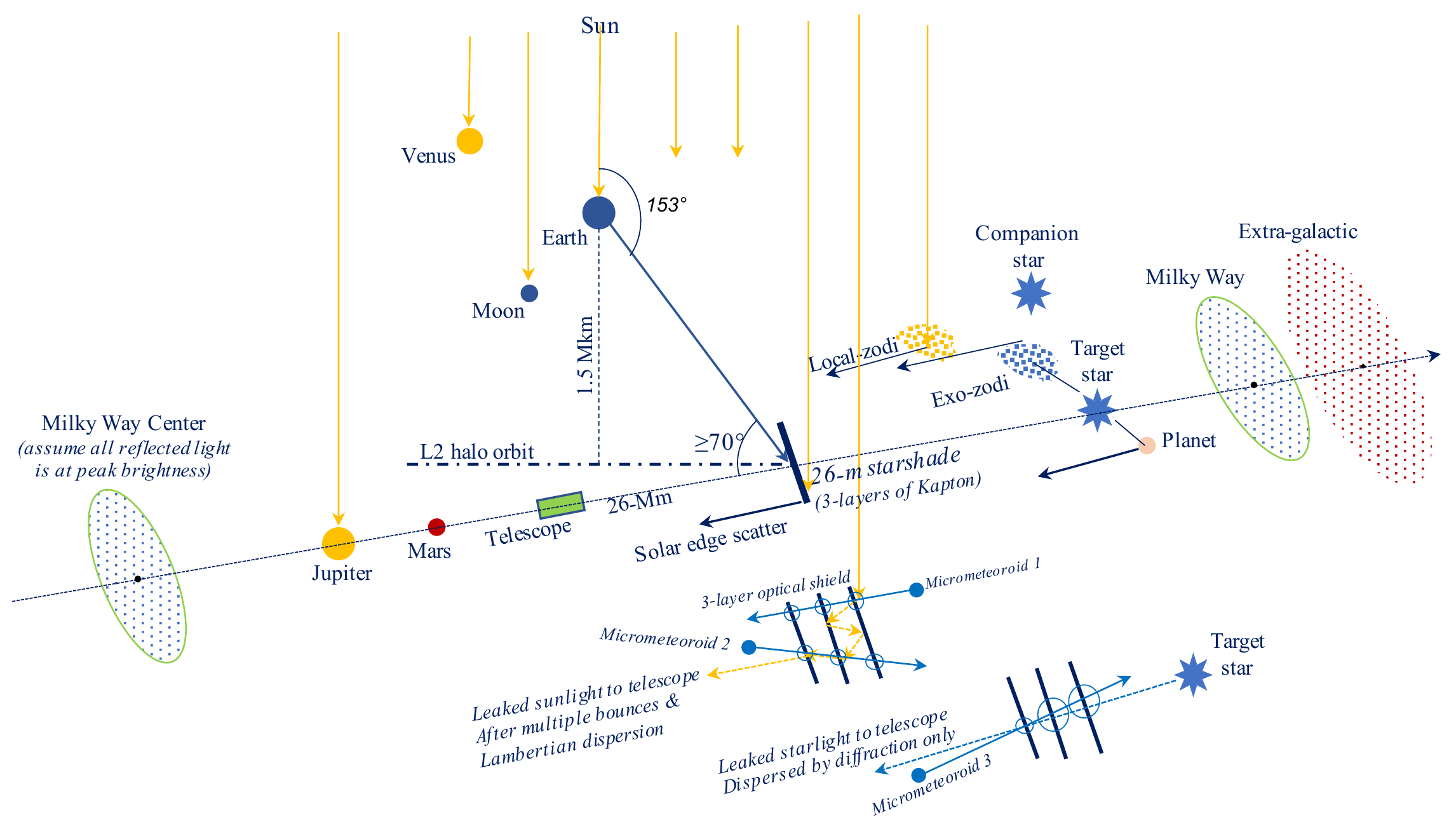}
\caption{The geometry of starshade-enabled exoplanet imaging and overview of stray light sources. Sunlight scattered to the telescope by the starshade's optical edge (i.e., the solar glint) is the dominant stray light source, followed by the reflection of the Milky Way, Earth, and other bright bodies in the Solar System. Micrometeoroids can produce holes on the starshade's optical shield and cause leakage of sunlight and starlight. In exoplanetary systems, the exozodiacal dust can scatter the host star's light to the telescope. Finally, other more distant stars and galaxies may appear on the image and cause confusion. The quantities specified are for Starshade Rendezvous with {\it Roman} but HabEx would be qualitatively similar. We provide a comprehensive analysis of these light sources and their impact on the science performance in Sections~\ref{sec:model}--\ref{sec:analysis}.
}
\label{fig:schematic}
\end{figure}

Fig. \ref{fig:schematic} provides an overview of the geometry of starshade-enabled exoplanet imaging and an overview of the background and noise sources. Regardless of the specifics of telescopes, the S/N of exoplanet direct imaging with a starshade is
\begin{equation}
    {\rm S/N} = \frac{N_{\rm P}}{\sqrt{N_{\rm P}+\alpha(N_{\rm S}C+N_{\rm G}+N_{\rm E}+N_{\rm Z}+N_{\rm D})+\beta^2(N_{\rm S}C+N_{\rm G}+N_{\rm E}+N_{\rm Z}+N_{\rm D})^2}}, \label{eq:sn}
\end{equation}
where $N_{\rm P}$, $N_{\rm S}$, $N_{\rm G}$, $N_{\rm E}$, $N_{\rm Z}$, and $N_{\rm D}$ are counts from the planet, the star, the solar glint, the exozodiacal dust, the local zodiacal dust, and detector noise, and C is the instrument contrast. The parameters $\alpha$ and $\beta$ in Eq. (\ref{eq:sn}) result from background subtraction, which will be discussed in Section \ref{sec:background}. The stray light sources shown in Fig. \ref{fig:schematic} but not included in Eq. (\ref{eq:sn}) will be discussed in Section~\ref{sec:stray}. The counts are defined by a photometric aperture, which in turns corresponds to the point spread function (PSF) of the telescope. We assume a photometric aperture diameter of $\lambda/D$, which is consistent with the S5 milestone reports\cite{Harness2019a,Harness2019b,Hilgemann2019}. Our choice of photometric aperture would encircle 46\% of the flux from a point source, and is smaller than Ref. \cite{Stark2014,stark2019exoearth}, which is $1.4\lambda/D$. The S/N yielded from the $\lambda/D$ photometric diameter is approximately 7\% less than the theoretical maximum achieved at $1.4\lambda/D$ in the background-limited regime. The aperture size in reality may eventually be controlled by the fixed pixel scale of the detector. Parameters used to estimate the contribution of the background sources and the expected S/N are summarized in Table \ref{tab:parameter}.

\subsection{Background Removal}
\label{sec:background}

The quality of background subtraction is characterized by the parameters $\alpha$ and $\beta$ in Eq. (\ref{eq:sn}). When $\beta=0$, the background can be subtracted to the photon-noise limit, where $\alpha$ describes the fractional increase from the photon noise that would manifest in the result. For example, if image processing can be approximated by subtracting two adjacent pixels of equal background, one with the planet and the other without, $\alpha=2$. For another example, if many more pixels can be used to characterize the background, $\alpha$ would approach unity. One may thus reasonably expect that with largely smooth and static background, $\beta\rightarrow0$ and $\alpha\sim1-2$. Note that all counts in Eq. (\ref{eq:sn}) are proportional to the integration time ($\Delta T$). When $\beta\rightarrow0$, ${\rm S/N}\propto\Delta T^{-1/2}$.

Imperfect background removal would lead to a small but nonzero $\beta$. Let $K$ be $N_{\rm P}/N_{\rm B}$, where $N_{\rm P}$ is the count from the planet and $N_{\rm B}$ is the count from the dominant background or noise term. Eq. (\ref{eq:sn}) can be converted to the following form:
\begin{equation}
{\rm S/N} = \frac{\rm S/N_0}{\sqrt{1+\frac{\beta^2}{K^2}({\rm S/N_0})^2}}, \label{eq:sn1}
\end{equation}
where ${\rm S/N_0}$ is the S/N when $\beta=0$. Eq. (\ref{eq:sn1}) expresses how much S/N would degrade due to imperfect background calibration. Several insights can be observed from Eq. (\ref{eq:sn1}). First, when $K\gg1$ (i.e., the planet dominates over the background), the S/N is not prone to degradation due to imperfect background calibration. Second, when $K\ll1$ (i.e., the background dominates over the planet), or at large ${\rm S/N_0}$ (i.e., large $\Delta T$), the asymptotic S/N would be ${\rm S/N}\rightarrow K/\beta= N_{\rm P}/\beta N_{\rm B}$. Third, at the critical case of $\beta=\frac{K}{{\rm S/N_0}}$, the S/N would become ${\rm S/N_0}/\sqrt{2}$, i.e., degraded by a factor of $\sqrt{2}$. Note that the value of $\beta$ can be different for each source of background; as we will show later, one source would usually dominate for each observation. Here we neglect potential interference between the noise sources and its potential contribution to the $\beta$ term.

The analysis here tells us that the fundamental limit of planet detection is determined by the flux ratio between the planet and the dominant background source ($K$ in Eq. \ref{eq:sn1}), and how well image processing can subtract the background to the photon-noise limit ($\beta$ in Eq. \ref{eq:sn1}). Causes for not achieving the photon-noise limit include detector systematics (as is the case for transit observations with Hubble and Spitzer\cite{greene2016characterizing}) and temporal variability of speckles in coronagraphic direct imaging\cite{krist2018wfirst}. These causes do not apply to future starshade direct imaging because of the use of EMCCD and the decoupling between starlight suppression from telescope optics. The capability of deep imaging provided by starshade may however render other causes to be the limiting factor, for example, the temporal variability of residual starlight and solar glint and the spatial distribution of exozodiacal dust. We adopt $\alpha=2$ and $\beta=0$ in the analyses that follow, and come back to discuss this point in Section \ref{sec:discussion}.

\begin{figure}[!h]
\centering
\includegraphics[width=0.6\textwidth]{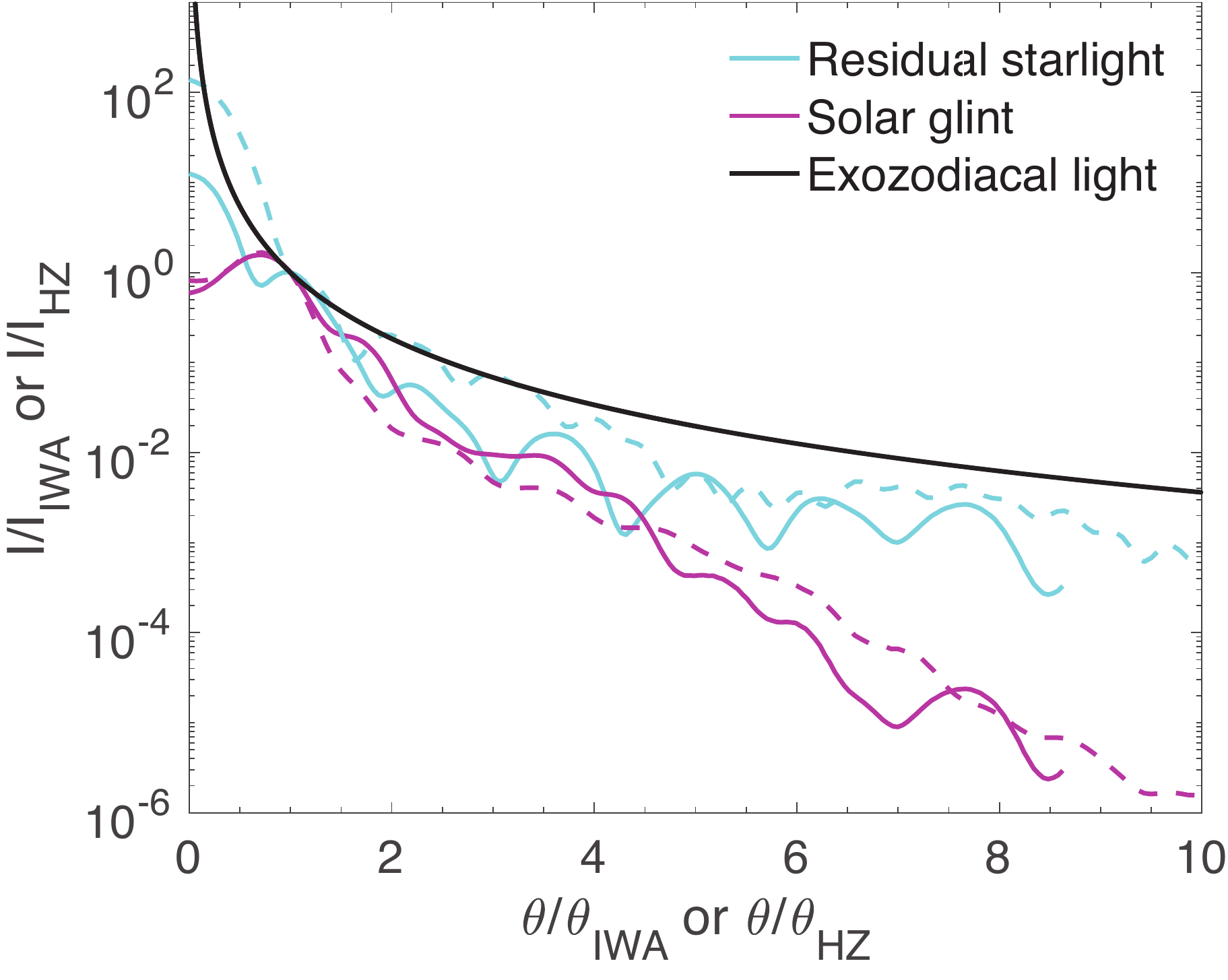}
\caption{Dependencies of residual starlight, solar glint, and exozodiacal light on angular separation. The solid lines are for a 26-m starshade coupled with {\it Roman} and the dashed lines are for a 52-m starshade of HabEx. The residual starlight and solar glint shown are azimuthal averages and scaled to the IWA, and their dependencies on the off-axis angle $\theta$ are calculated using SISTER. For \textit{Roman}, the IWA in the green band (615--800 nm) is 104 mas, corresponding to a 13-m radius at a distance of 26 Mm. For Habex, the IWA is 70 mas. This scaling conveniently lets one describe the residual starlight and the solar glint, the source of which is always located within the geometrical starshade pattern, in terms of apparent starshade radius. The exozodiacal light is scaled to the habitable zone (HZ), and its angular dependency is calculated with Zodipic\cite{kuchner2012zodipic} assuming a Solar-System dust density profile.}
\label{fig:scaling}
\end{figure}

\begin{table}[!h]
    \centering
    \begin{tabular}{lcc}
    \hline\hline
    Parameter & {\it Roman} & HabEx \\
    \hline
    \multicolumn{3}{l}{\it Starshade} \\
    \hline
    Diameter & 26 m & 52 m  \\
    Distance to telescope & 25.7 Mm & 76.6 Mm  \\
    IWA & 104 mas & 70 mas \\
    Contrast at IWA & $10^{-10}$ & $10^{-10}$ \\
    Solar glint magnitude$^*$ at IWA & 24.6 & 27.2 \\
    & (615-800 nm) & (300-1000 nm) \\
    Solar glint magnitude$^*$ with coating & 27.1 & 29.7 \\
    \hline
    \multicolumn{3}{l}{\it Telescope} \\
    \hline
    Aperture & 2.4 m & 4.0 m \\
    End-to-end throughput & 0.03 & 0.2 \\
    Detector dark current & $3\times10^{-5}$ e/pix/s & $3\times10^{-5}$ e/pix/s \\
    Detector clock-induced charge & $1.3\times10^{-3}$ e/pix/frame & $1.3\times10^{-3}$ e/pix/frame \\
    Pixels per spectral element & 42 & 56 \\
    \hline
    \multicolumn{3}{l}{\it Observation} \\
    \hline
    Wavelength & 700 nm & 700 nm \\
    Spectral resolution & 70 & 140 \\
    \hline
    \multicolumn{3}{l}{\it Planet} \\
    \hline
    Geometric albedo & 0.3 & 0.3 \\
    Phase angle & 60$^{\circ}$ & 60$^{\circ}$ \\
    Phase function & Lambertian & Lambertian \\
    Exozodi dust level & 3 zodis & 3 zodis \\
    \hline
    \hline
    \end{tabular}
    \caption{Parameters adopted in this work. *The magnitude is defined such that a 25-magnitude glint would have the same energy flux in the wavelength band in parenthesis as a solar-spectrum point source that has a magnitude of 25 relative to an A0V star, analog to the Johnson magnitude system.}
    \label{tab:parameter}
\end{table}

\subsection{Residual Starlight}
\label{sec:contrast}

Based on the S5 milestone reports\cite{Harness2019a,Harness2019b}, we adopt a contrast ratio of $10^{-10}$ at the IWA resulting from an imperfect starshade. As the habitable zones of nearby stars are often substantially larger than the IWA, we use the \textit{Starshade Imaging Simulation Toolkit for Exoplanet Reconnaissance} (SISTER\cite{Hildebrandt2020JATIS},  http://sister.caltech.edu) to determine the instrument contrast as a function of the angular separation (Fig. \ref{fig:scaling}) and use the results in the subsequent analyses. The residual starlight drops with the angular separation as approximately $\theta^{-3.4}$, where $\theta$ is the off-axis angle. This is slightly steeper than the expected Airy pattern drop of $\theta^{-3}$ for a filled aperture, due to the distribution of the residual starlight (i.e., some near the center of the starshade and some localized near the petals). Lastly, an off-axis companion beyond the IWA does not experience a change in transmission for starshade observations\cite{Hildebrandt2020JATIS}.

\subsection{Solar Glint}
\label{sec:glint}

While the sun will not be on the telescope-facing side of the starshade, the starshade's edges will scatter sunlight towards the telescope via a combination of diffraction, diffuse reflection, and specular reflection\cite{martin2013starshade}. When these scattering mechanisms are considered together, the telescope will see the scattered sunlight coming mainly from localized regions on a few petals where the optical edge is aligned for specular reflection (Fig. \ref{fig:schematic}). These will appear as two broad lobes due to the telescope’s finite spatial resolution, i.e., the solar glint. The solar glint is a unique effect in direct imaging using a starshade.

We adopt the brightness of the solar glint measured by S5 with a razor-sharp, amorphous metal edge\cite{Hilgemann2019}, and use SISTER to calculate its angular dependency (Fig. \ref{fig:scaling}). The expected magnitudes of the solar glint are calculated by combining scattering measurements of prototype optical edge segments and the optical models of the starshade\cite{Hilgemann2019}. We adopt a ``worst case'' scenario, i.e., the maximum solar angle ($\sim83^{\circ}$) and 95\% confidence upper limit of the brightness at the IWA\cite{Hilgemann2019}. The expected brightness of the solar glint for HabEx is approximately one-order-of-magnitude less than that for {\it Roman}. This is because the starshade of HabEx would be much more separated from the telescope than the starshade from {\it Roman} (Table \ref{tab:parameter}). The solar glint features a hump that peaks at the IWA and quickly drops as the angular separation increases, and thus impacts planet search near the IWA most significantly.

Recognizing the potential impact of the solar glint on science performance, S5 and its Science and Industry Partnership have been actively seeking improvement on the optical edge technology. A multi-layer, thin-film coating has recently emerged as a highly promising design, and experiments have indicated that the coating would result in a solar glint brightness lower than the uncoated design by approximately one order of magnitude\cite{McKeithen2020JATIS}. We will also study the science performance with the solar glint brightness suppressed by the coating.

%imilarly from the S5 milestone report\cite{Hilgemann2019} and example SISTER simulations, we approximate the dependency of the solar glint on the angular separation as $\propto\theta^{-4.87}$. The solar glint drops to zero when $\theta$ is more than $\sim6$ times the inner working angle. Compared with residual starlight, the solar glint decreases with angular separation more quickly. 
%The magnitudes of the solar glint are reported at two wavelengths ($\sim500$ and 700 nm), and we interpolate linearly for the magnitude at other wavelengths. 
%We adopt a ``worst case'' scenario, i.e., the maximum solar angle ($\sim83^{\circ}$) and 95\% confidence upper limit of the brightness at the IWA\cite{Hilgemann2019}. 

%Note that the expected brightness of the solar glint for HabEx is approximately one-order-of-magnitude less than that for {\it Roman}. This is because the starshade of HabEx would be much more separated from the telescope than the starshade rendezvous from {\it Roman} (Table \ref{tab:parameter}). 

%Generally, the solar glint is brighter at longer wavelength because the starshade would be closer to the telescope.

\subsection{Exozodiacal Light}
\label{sec:exozodi}

Results from the most sensitive exozodiacal dust survey indicate that the majority of nearby Sun-like stars have relatively low habitable-zone dust levels, with the best-fit median to be 3 times the Solar-System level, while some stars (e.g., $\epsilon$ Eridani) are significantly more dusty\cite{Ertel2020}. In this work, we assume ``1 zodi'' corresponds to 22 mag arcsec$^{-2}$ in the V band at the habitable zone\cite{Stark2014} and typically assume ``3 zodis'' in the analysis. This assumption was also adopted by mission concept studies and exoplanet yield analyses\cite{stark2019exoearth,Seager2019,Gaudi2020}. We discuss the sensitivity of the exozodi levels in Section~\ref{sec:discusszodi}. We also assume that the brightness of the exozodiacal light is independent of the distance or the stellar type if evaluated at the habitable zone\footnote{Unless otherwise noted, the habitable zone in this paper refers to the orbital distance that would yield the same stellar flux as 1 AU from the Sun, i.e., the 1-AU equivalent.}, besides a factor from the spectral shape of the star\cite{Stark2014}. Scaling to the habitable zone, we approximate the brightness of the exozodiacal light as dependent on the semi-major axis as $a^{-2.44}$. This scaling is based on example ``Zodipic'' simulations with a Solar-System dust density profile\cite{kuchner2012zodipic}. 

\subsection{Detector Noise}
\label{sec:detector}

We model the detector noise as the combination of dark current and clock-induced charge\cite{stark2019exoearth}. The EMCCD would have effectively zero read noise. We assume the frame rate to be 6.73 times the count rate of the brightest pixel\cite{stark2019exoearth}. This is to ensure that the probability of two photons arriving at the brightest pixel to be less than 1\%, at the expense of increasing the clock-induced charge. The number of pixels for each spectral element is assumed to be 42 for {\it Roman} and 56 for HabEx. These are estimated for spectral characterization at 700 nm with the assumptions of a PSF core of 4 pixels, dispersed into 6 pixels per spectral elements, and the detector providing Nyquist sampling at 400 nm for {\it Roman} and 300 nm for HabEx\cite{stark2019exoearth}.

%The detector noise is dominated by dark current and is $\sim10$ counts per hour.

%{\bf TBD from Stefan} To clarify a little about the QE for EMCCDs in high gain mode: the oft quoted 50\% reduction in QE is just a rule of thumb enabling an approximate SNR calculation. In reality, the QE of the device is still the QE of the device, but applying a reduction in QE of 50\% increases the effective noise. This kludge is applied to account for the multiplication noise (akin to “excess noise” in avalanche photodiodes). The multiplication noise arises because of the statistical variability in the overall gain; even with a 200 stage multiplier, because of the low probability of multiplication at any stage, a photoelectron can come through without being multiplied or come through having been multiplied many times. This spread of outcomes is the multiplication noise.

\section{Overview of Stray Lights}
\label{sec:stray}

Here we overview other noise terms not accounted for in the primary noise budget (Eq. \ref{eq:sn}).

\subsection{Astrophysical Background}
\label{sec:astrobackground}

The very dark shadow created by the starshade will reveal not only planets but also faint stars and distant galaxies. The Exo-S final report\cite{seager2015exo} provided a detailed analysis of background star and galaxy confusion and suggested mitigation strategies. If planets within 5 AU from the parent stars are potentially detectable (farther planets would often be too faint in reflected light), the largest area for planet search would be $\sim6$ arcsec$^2$. One can expect $\sim2$ distant galaxies down to V$\sim31$\cite{illingworth2013hst} and $\sim0.2$ stars in this area depending on the galactic latitude\cite{seager2015exo}. Stars with known companions that have small angular separation and would impact direct imaging have been excluded from the target lists of Starshade Rendezvous\cite{Seager2019} with {\it Roman} and HabEx\cite{Gaudi2020}. Spectra may provide clues to tell the planets apart from these background sources, and revisits and the detection of common proper motions with the parent star would be required to confirm the planets.

\subsection{Micrometeoroid Holes}
\label{sec:micrometeo}

Micrometeoroids can penetrate the starshade’s multi-layer opaque optical shield (OS). The baseline OS design consists of three evenly spaced layers of Black Kapton. Some large or high-velocity micrometeoroids can produce through-holes on the starshade. During a science integration, on-axis starlight can pass directly through the fraction of particle trajectories aligned to the starshade normal to disperse only via diffraction towards the telescope (Fig. \ref{fig:schematic}). Off-axis sunlight instead requires multiple reflections within the OS before exiting to disperse in the Lambertian fashion towards the telescope (Fig. \ref{fig:schematic}). It is thus necessary to consider micrometeoroid holes and the transmission of starlight and sunlight through them.

We have developed a model to estimate micrometeoroid holes and the resulting brightness levels for Starshade Rendezvous with {\it Roman}. The mass-flux distribution of micrometeoroids at L2 is estimated with the Grun model\cite{grun1980orbital}. The Grun model does not account for seasonal meteor showers that bring elevated fluxes by 1 -- 2 orders of magnitude. It may be necessary to orient the starshade to a near edge-on direction during one or two showers per year, for a total of approximately one month per year. The lost observation time can be mitigated to some extent by the timing of retargeting maneuvers with long coast periods.

The minimum particle mass required to enter the OS is computed with a single layer ballistic equation from NASA's Preferred Reliability Practices document for micrometeorite protection (Standard PD-EC-1107). The minimum particle mass to pass the middle layer and then exit the third layer are computed with a two-layer ballistic equation\cite{cour1979space}. The exit computation conservatively neglects the benefit of the middle layer. The incoming flux is assumed isotropic in direction considering variable starshade pointing throughout the mission. The particle specific density is conservatively bounded by a constant of 2.5 g cm$^{-3}$ for considering the ability to penetrate the OS and 1 g cm$^{-3}$ for considering the size of holes produced. Entry holes diameters approximately match the particle diameter for large particles and can be greater than the particle diameter by up to a factor of five for small particles\cite{horz2012cratering}. When a high-velocity particle penetrates the first layer, the particle and the shield material will vaporize, creating an expanding gas cloud. A subset of the gas cloud can then penetrate the middle and exit layers. We estimate that the exit and middle-layer hole diameters can grow by up to a factor of thirty from the original particle diameter to account for the gas cloud expansion\cite{schonberg1995debris}.

Our model indicates that the hole areas after three years on-orbit would be 0.1 parts-per-million (ppm) by surface area entry holes on both sides, 50 ppm exit holes on both sides, and 150 ppm middle-layer holes. The expected number of entry holes is about $5\times10^5$, while the expected number of exit holes is $<400$. The exit holes are produced only by large particles.

We estimate that $\sim10\%$ through-holes would provide a direct path for starlight to the telescope. The leakage is limited by the entry hole area on both sides to yield 0.02 ppm of effective area. Feeding this area to our optical performance model of starshades\cite{shaklan2010error}, we estimate that the starlight leakage due to micrometeoroid holes to correspond to a residual starlight contrast of $10^{-13}$, lower than the allowed residual starlight contrast by three orders of magnitude.

We then estimate the upper bound of solar transmission to be the product of the porosity factors of the layers and obtain $4\times10^{-13}$. The solar leakage is also proportional to micrometeoroid hole area, but not necessarily the number of holes. This transmission corresponds to a brightness of solar leakage at the telescope that has a visual magnitude of 39.6 at the IWA at 700 nm, with the brightness falling off with the off-axis angle at approximately the same rate as for the solar glint. We thus expect the solar leakage due to micrometeoroid holes to be dimmer than the solar glint by $>10$ magnitudes. 
% A more precise model of solar transmission is being developed to include the effect of diffraction at each hole. We plan to validate the transmission model in an optical testbed currently under development. 

While the estimates apply for Starshade Rendezvous with {\it Roman}, they can be scaled for HabEx considering a longer (5-year) mission lifetime and a higher angular resolution (defined by $\lambda/D$). The distance from the starshade to the telescope does not appear in this scaling because the solid angle per resolution element is independent of the distance. We estimate that the brightness from micrometeoroid holes on HabEx's starshade would have approximately the same magnitude as the Starshade Rendezvous, and remain much dimmer than the solar glint.

\subsection{Bright-body Reflections}
\label{sec:brightbody}

Bright celestial bodies positioned on the telescope side of the starshade may have a portion of their light reflected towards the telescope, with the brightness falling off with the off-axis angle at approximately the same rate as for solar glint. The starshade presents mostly Black Kapton to the telescope and we evaluate bright-body reflections with its BRDF data. To estimate the magnitude of bright-body reflections, we start from a Solar-System planet's absolute magnitude, and consider the worst-case orbit phasing to derive the brightness incident at the starshade. We then take into account the angular size of the starshade and its reflectivity in the direction of the telescope. Lastly, we distribute the brightness into each resolution element on the starshade, as the starshade would have nearly uniform brightness in this problem. Table~\ref{tab:brightcal} provides a detailed walk-through of our estimates of bright-body reflections for Solar-System bodies.

\begin{table}[!h]
    \centering
    \begin{tabular}{llccc}
    \hline\hline
    Parameter & Units & Earth & Venus & Jupiter \\
    \hline
    Absolute magnitude ($H$)$^*$ & mags & -3.99 & -4.38 & -9.40 \\
    \hspace{4mm} Sun distance & AU & 1 & 0.72 & 5.2 \\
    \hspace{4mm} Starshade (SS) distance & AU & 0.0112 & 0.70 & 4.55 \\
    \hspace{4mm} Maximum phase angle at SS & deg & 153 & 90 & 9.3 \\
    \hspace{4mm} Phase function in SS direction & $\Delta$mags & 4.65 & 1.43 & 0.05 \\
    Brightness incident at SS & mags & -9.10 & -4.51 & -2.57 \\
    \hspace{4mm} Incidence angle from SS normal & deg & 70 & 50 & 0 \\
    \hspace{4mm} SS reflectivity in telescope direction & per s.r. & 0.00261 & 0.1 & 0.3 \\
    \hspace{4mm} SS area/distance$^2$ & $\Delta$mags & 30.8 & 30.8 & 30.8 \\
    Total brightness at telescope & mags & 29.2 & 29.1 & 29.4 \\
    \hspace{4mm} \# resolution elements on SS &  & 10 & 10 & 10 \\
    Brightness per resolution element & mags & 31.7 & 31.6 & 31.9 \\
    \hline\hline
    \end{tabular}
    \caption{Estimates of Earth, Venus, and Jupiter brightness reflected by the starshade to the telescope at their orbital positions and the starshade's position and pointing that maximize these stray light sources. The estimates use the parameters of Starshade Rendezvous with {\it Roman}, and the result for HabEx would scale by the solid angle per resolution element. *The absolute magnitude of a Solar-System object is defined as the apparent magnitude that the object would have if it were 1 AU from both the Sun and the observer, and in conditions of ideal solar opposition\cite{mallama2018computing}.}
    \label{tab:brightcal}
\end{table}

Light from Earth, Moon, and Venus can be reflected towards the telescope at grazing incidence angles, but only at the extremes of starshade's orbit position and pointing (Fig. \ref{fig:schematic}). Earth and Venus maximally can appear at about the same magnitudes of 31.7 mags at the IWA (Table \ref{tab:brightcal}). To have this magnitude, Venus must be at the quadrature orbital phase viewed from the starshade when the starshade is at one extreme of its L2 orbit, a combination that would rarely occur. The Moon will be much dimmer than Earth and Venus for starshade reflection to the telescope.

Light from Mars and Jupiter can be reflected towards the telescope at nearly normal incidence angles (Fig. \ref{fig:schematic}). We find that Jupiter will never exceed a brightness of 31.9 mags at the IWA (Table \ref{tab:brightcal}), and Mars will be much dimmer.

In addition to the Solar-System bodies, the integrated light from the center of the Milky Way will likely be the brightest object that can be reflected towards the telescope by the starshade. The Milky Way can appear as bright as 20.6 visual magnitudes per arcsec$^{-2}$\cite{hoffmann1998photographic}, or 27.0 mags per resolution element at 700 nm. The starshade's hemispherical reflectance is about 5\%. Assuming that all of the reflected light comes from the brightest part of the Milky Way and adjusting for the starshade not being a full disk, we estimate the maximum brightness to be 31.3 mags at the IWA.

In all, we find that the bright-body reflections combined would be no brighter than a magnitude of $\sim30$ at the IWA as a conservative estimate for Starshade Rendezvous with {\it Roman}. The bright-body reflections for HabEx should be dimmer by a factor of $\sim2$. The bright-body reflections thus constitute a stray light source dimmer than the solar glint (with coating improvement) by approximately one order of magnitude for Starshade Rendezvous and a factor of 2 for HabEx.

\subsection{Other Stray Light}
\label{sec:otherstray}

Other stray light may include secondary solar reflections, fluorescence, thruster exhaust solar scatter, and stray light produced by the telescope. The first three potential sources are subjects of starshade technology and our preliminary analyses indicate that these light sources likely contribute to the background by no more than a fraction of the solar glint. S5 is conducting extensive stray light analyses to evaluate possible stray light paths and, if necessary, adjust the starshade design and the observation constraints to mitigate their impacts.

The current starshade design is intended to preclude any sunlight reaching the telescope after only a single reflection, except for the optical edges (i.e., the solar glint). 
%{\bf (what is this?) One exception is the gap between optical edge segments that are about 1-m long and this is under study. It is understood that these gaps need to be filled to some extent, after installation on petals, and one goal of the stray light analysis effort is to specify that extent.}
However, secondary solar reflections, where sunlight can reach the telescope after two reflections, are possible with certain particular light paths and they are being studied by S5. Preliminary analysis suggests an acceptable brightness level. For example, some stray light paths involve an out-of-plane deformed petal and stray light analyses are being developed to establish a tighter, yet still readily achievable requirement of how far the petal can deform out-of-plane versus the current specification based on diffraction performance alone. For another example, some solar leakage may occur through small gaps between the optical shield at the inner disk to petal interface. This interface is dynamic at the end of disk deployment as a hard tie-down point rotates into node plates at the ends of each inner disk facet. A labyrinth seal is designed to attenuate the sunlight with a preliminary verification of performance, and a more thorough stray light analysis is planned to confirm this design.

Black Kapton used in the starshade's OS is known to glow in response to high energy solar electrons due to material fluorescence. Informed by past studies for JWST, we estimate a relatively dim magnitude of 33.3 at the IWA, i.e., dimmer than the solar glint (with coating) by two orders of magnitude. Efforts are underway to better understand the input electron energy levels and evaluate the fluorescence of Black Kapton and other material refinements.
%There is some residual concern, however, for model uncertainty that results from a limit on the tested electron energy levels. An effort is underway to first better understand the original test data, second explore the existence of supplemental test data and, if necessary, consider additional tests and model validation efforts.
%Fluorescence can also be mitigated by coatings or other material refinements. A dark coating for Black Kapton is already under study and fluorescence testing is called for. Another possibility is to develop a customized version of Black Kapton, a 3M product. The carbon mixed into the Kapton is conductive and JWST test images suggest discrete glowing regions that could be where carbon particles are less dense. A hypothetical material refinement could be to specify a more uniform distribution of carbon. More generally, a broader future effort will be required to evaluate the fluorescence of all starshade materials in view of the telescope. This will extend into the flight program as material choices mature, including on the spacecraft bus system.

Periodic thruster firings to maintain formation flight are expected about once every 10 minutes\cite{Flinois2018}. Exhaust particles will scatter sunlight at a level that can saturate detectors. An operating concept to prevent detector saturation is for the starshade to notify the telescope of an imminent thruster firing and for the telescope to switch to a fast detector read-out mode for some conservative duration (e.g., 10 seconds). This is more than enough time for the bulk of the exhaust to leave the field of view and only causes a modest loss of observing time. Lastly, large exhaust particles might ``loiter'' around the starshade or stick to the exhaust nozzle to be pushed off at relatively low velocity at the next thruster firing. The optical density of these loitering particles would be much lower than the high-velocity exhaust particles, and detailed simulations are required to assess their brightness.

%The direct transmission of sunlight through the optical shield must be limited by material opacity. The carbon filled Black Kapton is considered highly opaque based upon early measurements, but more precise measurements to confirm this are planned using the Princeton optical testbed used to verify optical performance.

% Micrometeoroids will create light flashes as particles vaporize at impact. An impact event that produces just one photon that enters the telescope barrel is estimated to occur about once per hour. Impact events that produce enough photons to register above the noise floor are expected to be rare.

%Straylight direct to the telescope also needs to be considered. Observing constraints will exist for bright objects in front of the telescope, including the Milky Way center.

\section{RESULTS AND ANALYSES}
\label{sec:analysis}

\begin{figure}[!h]
\centering
\includegraphics[width=0.6\textwidth]{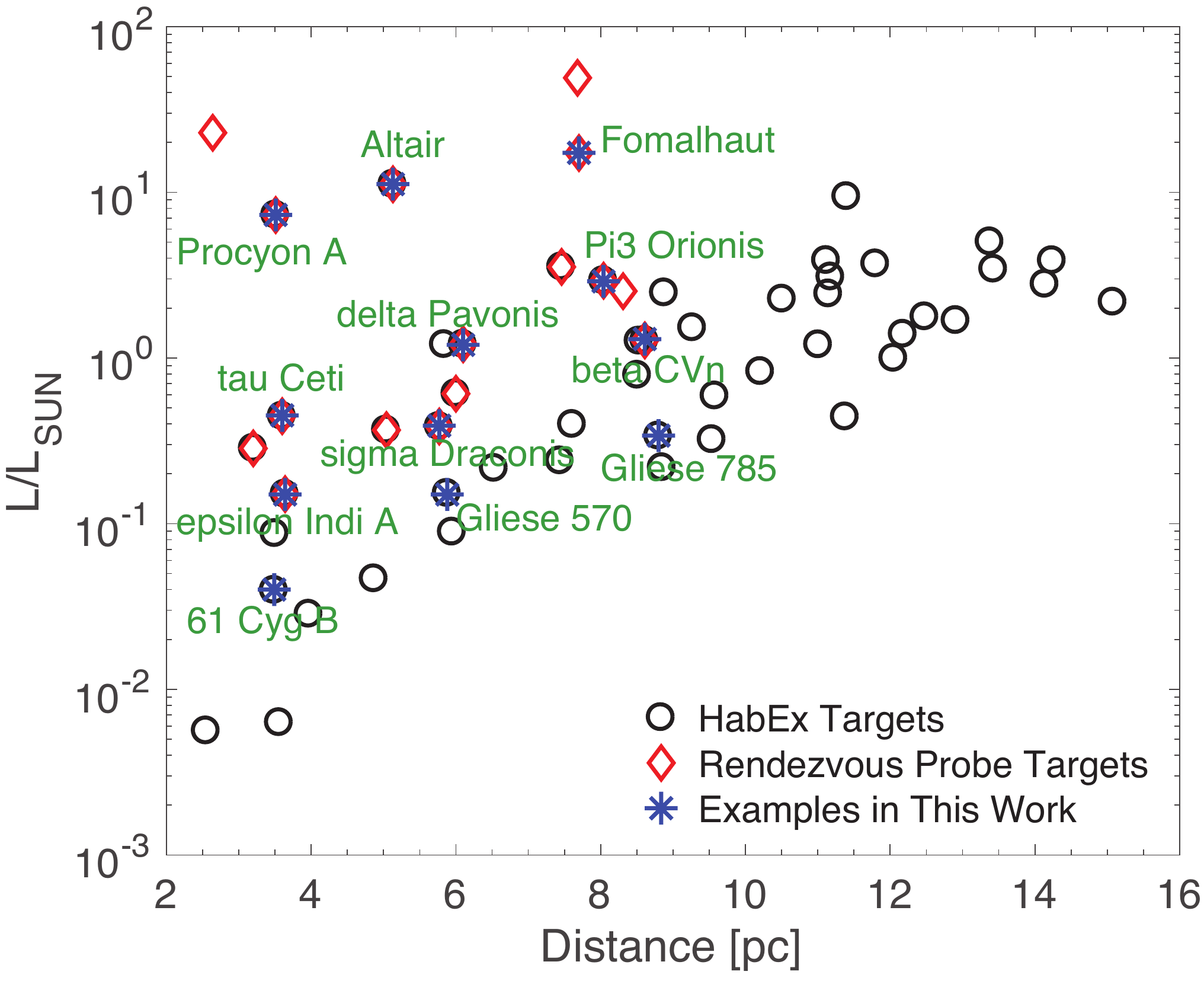}
\caption{Nearby stars for the search of potentially habitable planets adopted by Starshade Rendezvous\cite{Seager2019} and HabEx\cite{Gaudi2020}. Example stars used in this study are labeled. While many more M type stars can be found in the distance range shown, their habitable zones have smaller angular separations than the IWA of Starshade Rendezvous with \textit{Roman} or HabEx.}
\label{fig:target}
\end{figure}

We choose stars that represent the target lists of Starshade Rendezvous\cite{Seager2019} and HabEx\cite{Gaudi2020} in this study. As seen in Fig. \ref{fig:target}, the nearest stars for the search of potentially habitable planets may be grouped into three distance groups: $<4$ parsecs, 5 -- 6 parsecs, and $\sim8$ parsecs. Within each group, Starshade Rendezvous with {\it Roman} would search for habitable-zone planets around stars from A/F to late G/early K spectral types. HabEx would extend the search to stars of late K and even early M spectral types, in addition to many more distant stars. Most nearby stars are M type stars, and they are generally not amenable for direct imaging of planets in the habitable zone due to IWA restrictions. The example stars used in this study cover the distance groups and the spectral type (and luminosity) ranges of the target stars of Starshade Rendezvous with {\it Roman} and HabEx.

For each representative star, we evaluate the detectability of planets of varied size and planet-star separation. Rather than focusing on 1-R$_{\oplus}$ planets that receive Earth-like insolation, we include larger planets at colder or hotter orbits as potential search targets. This is motivated by the diversity of planets revealed by current exoplanet searches and improved understanding of the habitable zones. {\it Kepler} planet surveys and planetary mass measurements available for a subset of the detected planets have indicated two populations of small planets\cite{fulton2018california}. The planets in the $<1.7$ R$_{\oplus}$ population are mostly rocky\cite{rogers2015most}, and the planets in the $1.7\sim3.0$ R$_{\oplus}$ population could be planets with H$_2$/He gas envelopes\cite{owen2017evaporation,jin2018compositional} and/or substantial water layers\cite{zeng2019growth,mousis2020irradiated}. The larger planets, if they have moderate-size atmospheres, can host liquid-water oceans and thus be potentially habitable\cite{pierrehumbert2011hydrogen,madhusudhan2020interior}. The ``habitable zones'' for planets with H$_2$-dominated atmospheres can be substantially more separated from the parent stars than those for planets with N$_2$- and CO$_2$-dominated atmospheres\cite{pierrehumbert2011hydrogen,seager2013exoplanet}. 

We focus on the spectral characterization of exoplanets in this paper. We adopt a requirement of S/N=20 per spectral element, as this is a conservative estimate of the precision needed to measure atmospheric abundances from the reflected-light spectra and potentially distinguish the types of planets\cite{Lupu2016,Nayak2017,feng2018,Hu2019B2019ApJ...887..166H,damiano2020exorel,damiano2020multi}. The expected integration time to detect planets in the broadband will be substantially less than what is shown in this section. Also, all results shown in this section assume unbiased calibration of the background to the photon-noise limit. We will come back to discuss background calibration in Section~\ref{sec:discussion}.

\subsection{Current Performance of Starshade Rendezvous}
\label{sec:current}

\begin{figure}[!h]
\centering
\includegraphics[width=1.0\textwidth]{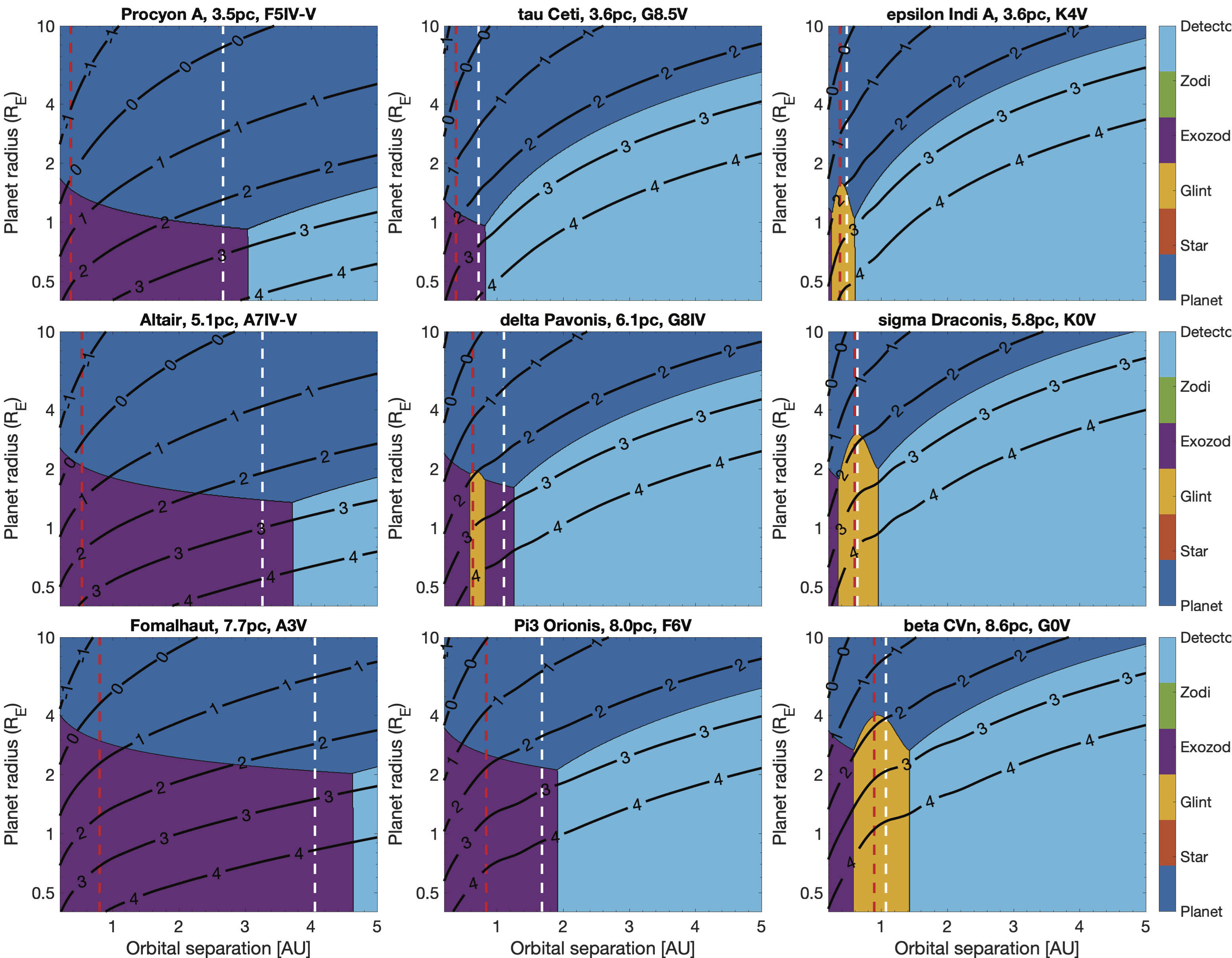}
\caption{Expected integration time (S/N=20 per spectral element, contour profiles, in base-10 log(hours)) and dominant noise source (shaded areas) expected in {\it Roman} and Starshade Rendezvous exploration of planets around nearby stars. The red dashed line corresponds to the IWA, and the blue dashed line corresponds to the planet-star separation for receiving Earth's insolation, i.e., the 1-AU equivalent. Parameters of the simulations are shown in Table \ref{tab:parameter}.}
\label{fig:noise}
\end{figure}

Fig. \ref{fig:noise} shows the expected integration time and the underlying dominant noise source to detect planets around nearby stars with demonstrated starshade optical performance and telescope parameters that approximately corresponds to the Starshade Rendezvous with {\it Roman} mission concept (see Table \ref{tab:parameter}). We make the following observations from Fig. \ref{fig:noise}. First, exozodiacal light is the dominant noise term for planet searches around nearby F and A stars. These early-type stars have widely separated habitable zones, and thus the habitable-zone planets would have small planet-to-star contrast. Also, because we assume the exozodiacal brightness scales to the habitable zone (the 1-AU equivalent in this context), the expected exozodiacal light brightness near the IWA is high for these early-type stars.

Second, for the search of small ($<2.5R_{\oplus}$) planets near the habitable zones of the nearest late G and K stars (i.e., $\tau$ Ceti and $\epsilon$ Indi A), the planet itself is likely the dominant noise term. In other words, the current performance expected for Starshade Rendezvous with {\it Roman} could provide a photon-limited detection, rather than background-limited detection, for temperate and small planets around these most favorable targets. This is because the stars are less luminous than the Sun, which makes the habitable zones closer to the stars (0.4 -- 0.7 AU). The planet-star contrast at their habitable zones would be 7 -- 20$\times10^{-10}$ for 1-$R_{\oplus}$ planets. Coupled with the brightness of these stars (visual magnitude of 3.5 -- 4.7), the brightness of 1-$R_{\oplus}$ planets in their habitable zones would have a visual magnitude of $\sim26.3$, i.e., much brighter than a ``mean'' scenario of 30. As such, the common belief that the detection would be mainly limited by exozodiacal light does not apply for these most favorable targets. They are also the prime targets from an integration time standpoint: to spectroscopically characterize (with S/N=20) a 1-$R_\oplus$ planet near the habitable zone only requires integration of a few hundred hours, and the integration time reduces to $\sim100$ hours for larger rocky planets with 1.5-$R_\oplus$ radius. Among these stars, $\epsilon$ Eridani has an especially bright dust disk that would probably prevent the search of small planets\cite{Ertel2020}. $\tau$~Ceti have two super-Earth-mass planets ($M\sin{i}\sim3.9M_{\oplus}$) near the habitable zone detected by radial-velocity measurements\cite{tuomi2013signals,feng2017color}, and an outer dust disk ($>6$ AU) detected by far-infrared and radio observations\cite{greaves2004debris,lawler2014debris,macgregor2016alma}. The constraints on the orbital elements and masses of these planets would aid future characterization.

Third, for slightly farther stars (5 -- 6 parsecs, e.g., 40 Eridani, $\delta$ Pavonis, 82 Eridani, $\sigma$ Draconis), the search for small planets near the habitable zones would be limited by exozodiacal light and solar glint. For the later-type stars whose habitable zones are close to the IWA, the search for habitable-zone planets are particularly affected by solar glint. With the current performance\cite{Hilgemann2019} and assuming calibration of the solar glint to the photon-noise limit, to characterize a 1-$R_\oplus$ planet near the habitable zone requires integration of a few thousand hours, i.e., likely infeasible for a realistic space mission. Characterizing a $1.5\sim2.5$-$R_\oplus$ planet, however, only requires an integration time of $100\sim1000$ hours. Large rocky planets and more volatile-rich planets could thus be studied with spectroscopy, and thus these ``super-Earths'' represent the near-term opportunity for the search of habitable worlds on these stars\cite{hu2019super}.

Finally, for stars $\sim8$ parsecs away, planet characterization near the habitable zones would limit to planets not smaller than $\sim2$ $R_\oplus$, and the integration time required for smaller planets would be well longer than 1000 hours. The solar glint continues to be the dominant noise term, except for the F and A stars.

\subsection{Impact of Future Development}
\label{sec:impact}

\begin{figure}[!h]
\centering
\includegraphics[width=1.0\textwidth]{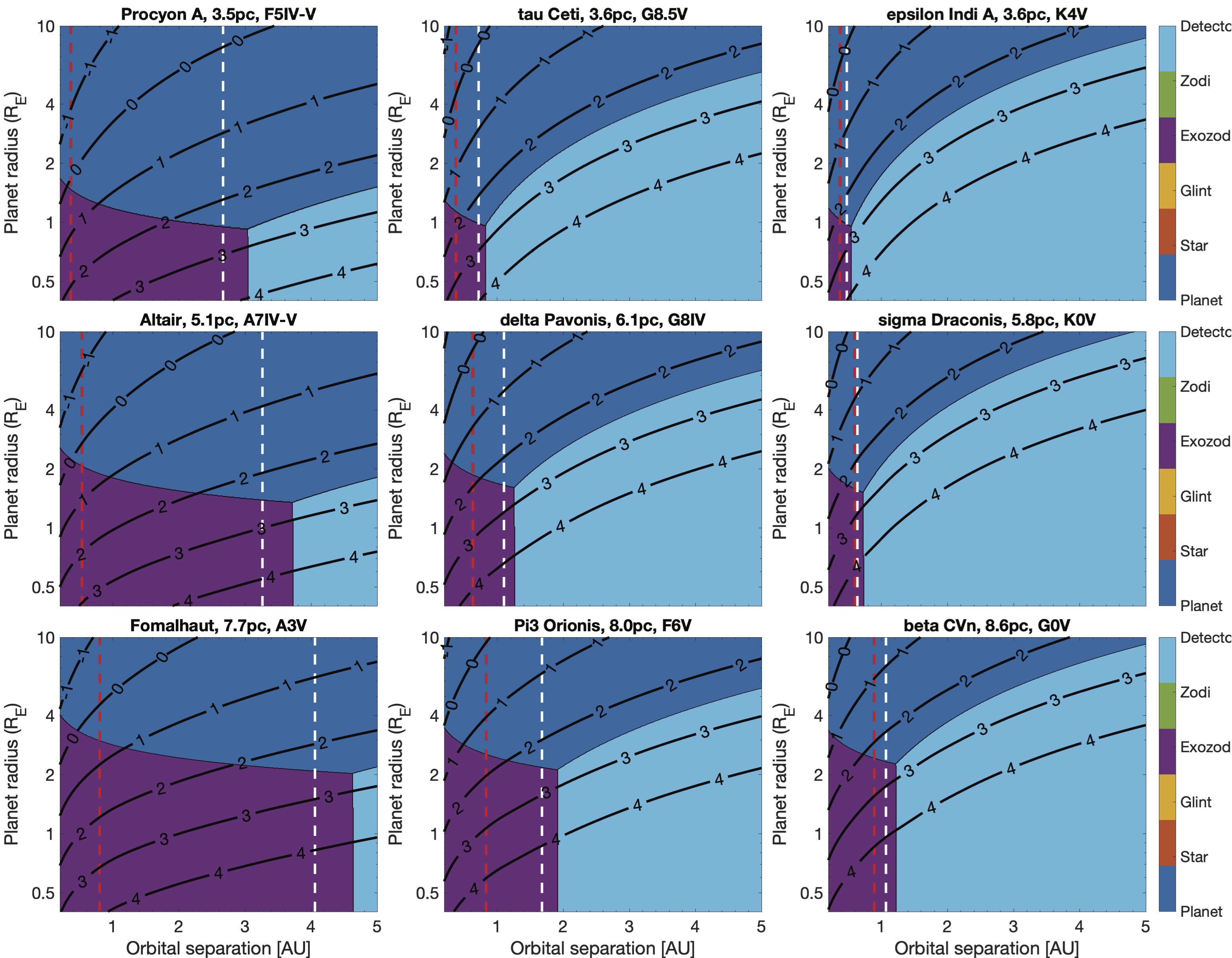}
\caption{The same as Fig. \ref{fig:noise}, except that the brightness of solar glint is lower than the nominal value by a factor of 10.}
\label{fig:noise_g}
\end{figure}

\begin{figure}[!h]
\centering
\includegraphics[width=0.6\textwidth]{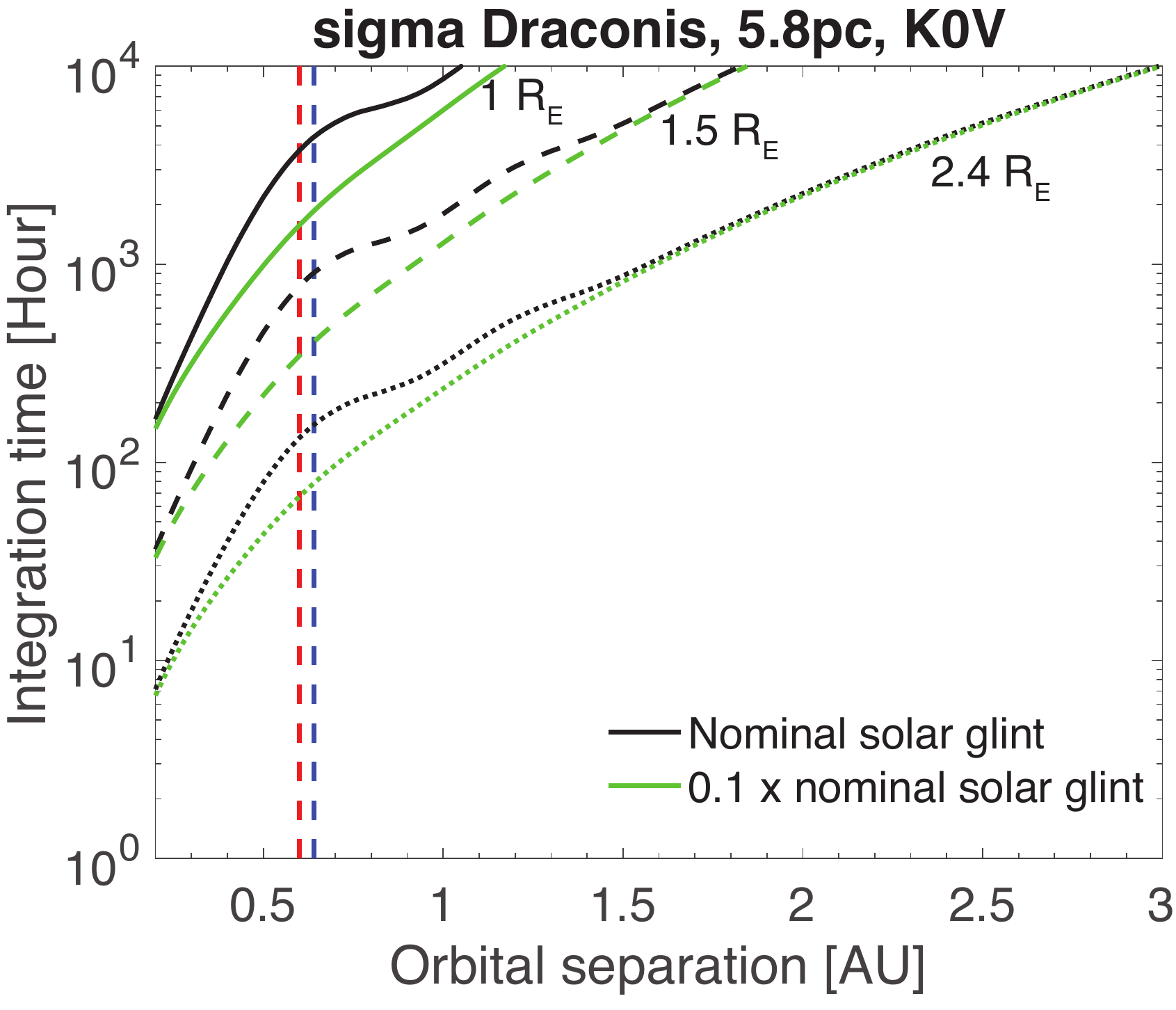}
\caption{Improvement in the required integration time to characterize (S/N=20 in a 10-nm wavelength channel) planets around a nearby star, from a 10-fold reduction of the solar glint brightness from the nominal value. This is essentially a zoom-in view of Fig. \ref{fig:noise_g} for sigma Draconis. The red dashed line corresponds to the IWA, and the blue dashed line corresponds to the planet-star separation for receiving Earth's insolation, i.e., the 1-AU equivalent.}
\label{fig:single_sg}
\end{figure}

\begin{figure}[!h]
\centering
\includegraphics[width=1.0\textwidth]{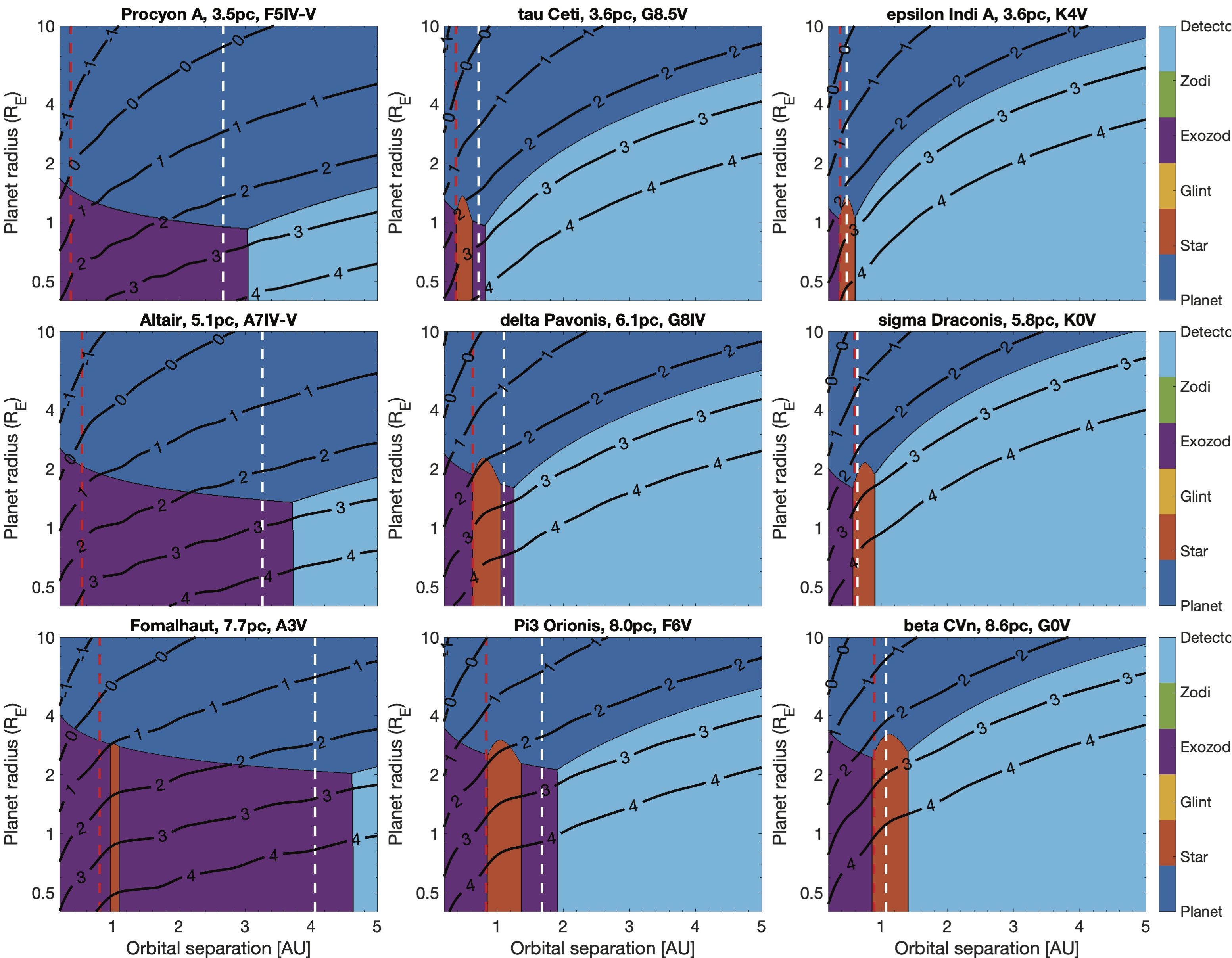}
\caption{The same as Fig. \ref{fig:noise}, except that the brightness of solar glint is lower than the nominal value by a factor of 10 and the residual starlight contrast is $3\times10^{-9}$ (i.e., higher than the nominal value by a factor of 30).}
\label{fig:noise_c}
\end{figure}

Guided by the analyses presented in Section~\ref{sec:current}, we now turn to the impact of the starshade performance parameters on the science returns. Because the diameter and the distance to the telescope determine the theoretical limits in starlight suppression and inner working angle\cite{glassman2009starshade}, we focus on the non-ideal effects. Specifically, random mechanical imperfections raise the starlight suppression level (expressed in the residual starlight contrast), and sources of stray light, dominated by the solar glint, contribute to the background.

As discussed in Section~\ref{sec:glint}, one may reasonably expect the actual solar glint brightness to be approximately one order of magnitude lower than the nominal value, due to coating technologies\cite{McKeithen2020JATIS}. Also, we have used a ``worst'' scenario in the analysis presented in Fig. \ref{fig:noise}, in that the Sun was at the maximum angle from the line of sight ($83^{\circ}$) and the planet was close to the position where the solar glint was maximized. An ``average'' scenario where the solar angle is less ($\sim53^{\circ}$) and the planet is away from the glint maximum would result in a reduction in the solar glint brightness by a factor of $\sim3$\cite{Hilgemann2019}. 

Figure \ref{fig:noise_g} shows the impact on the science performance from a 10-fold reduction on the solar glint brightness. We find that this improvement would drastically enhance the prospect of the search for small planets around later-type nearby stars. The solar glint would no longer be the dominant noise term for the planet observations around these stars. The exozodiacal light becomes the dominant noise term in place. Consequently, a decrease in the required integration time by a factor $2\sim3$ for planet search near the IWA would be expected (Fig. \ref{fig:single_sg}). This improvement is particularly substantial for exoplanet science, because this would reduce the time needed to characterize 1-$R_{\oplus}$ planets in the habitable zones of stars $5\sim6$ parsecs away from a few thousand to approximately 1000 hours (i.e., from infeasible to marginally feasible), and would reduce the time needed to characterize 1.5-$R_{\oplus}$ planets (i.e., large rocky planets) from $\sim1000$ hours to a few hundred hours (i.e., from marginally feasible to highly feasible). This improvement would also extend the search for planets around 8-parsec, Sun-like stars from 2-$R_\oplus$ planets to 1.5-$R_\oplus$ planets, i.e., into the regime of rocky planets. Together, these impacts would enlarge the stellar sample that would be amenable for searching rocky planets in the habitable zones by a factor of at least three, and thus a substantial improvement of the science prospect.

For starlight suppression, the nominal performance shown in Fig. \ref{fig:noise} does not have residual starlight as the dominant noise term anywhere in the explored parameter space. We are thus motivated to assess the tolerance of exoplanet characterization on a degraded residual starlight contrast. We have repeated the analysis with a contrast level that is 3, 10, 30, and 100 times higher than the $10^{-10}$ level demonstrated by S5 experiments, on top of the low-solar-glint scenarios shown in Fig. \ref{fig:noise_g}. We find no appreciable change in the search of temperate and small planets around nearby stars for up to 10 times worse contrast. Residual starlight would continue to be a non-dominating term in the noise budget, and the impact on the integration time would be minimal. Residual starlight would become the dominant noise term for observing 1-2 R$_{\oplus}$ planets of nearby stars, associated with a moderate increase in the integration time to achieve an S/N of 20, when the contrast is $3\times10^{-9}$ (i.e., 30 times worse, Fig. \ref{fig:noise_c}). This analysis indicates that the $10^{-10}$ starlight contrast is sufficient, with one order of magnitude margin, for the science performance of the Starshade Rendezvous.

To summarize, the demonstrated starshade optical performance coupled with {\it Roman} would enable spectral characterization of small and temperate planets of nearby stars. With the coating technology to reduce the solar glint, the characterization of Earth-sized and Earth-temperature planets will be possible for stars within $5\sim6$ parsecs, and the characterization of temperate and large rocky planets of 1.5 R$_{\oplus}$ will be possible for stars as far as $\sim8$ parsecs. A contrast level of $10^{-10}$ at the IWA is sufficient for these science applications, as the residual starlight appears nowhere in the parameter space as the dominant noise term. For a uniform dust level of 3 zodis, the dominant noise term for the small planet characterization is likely exozodiacal light. The impact of exozodiacal light is larger for farther stars.

\subsection{Expected Performance of HabEx}
\label{sec:habex}

\begin{figure}[!h]
\centering
\includegraphics[width=1.0\textwidth]{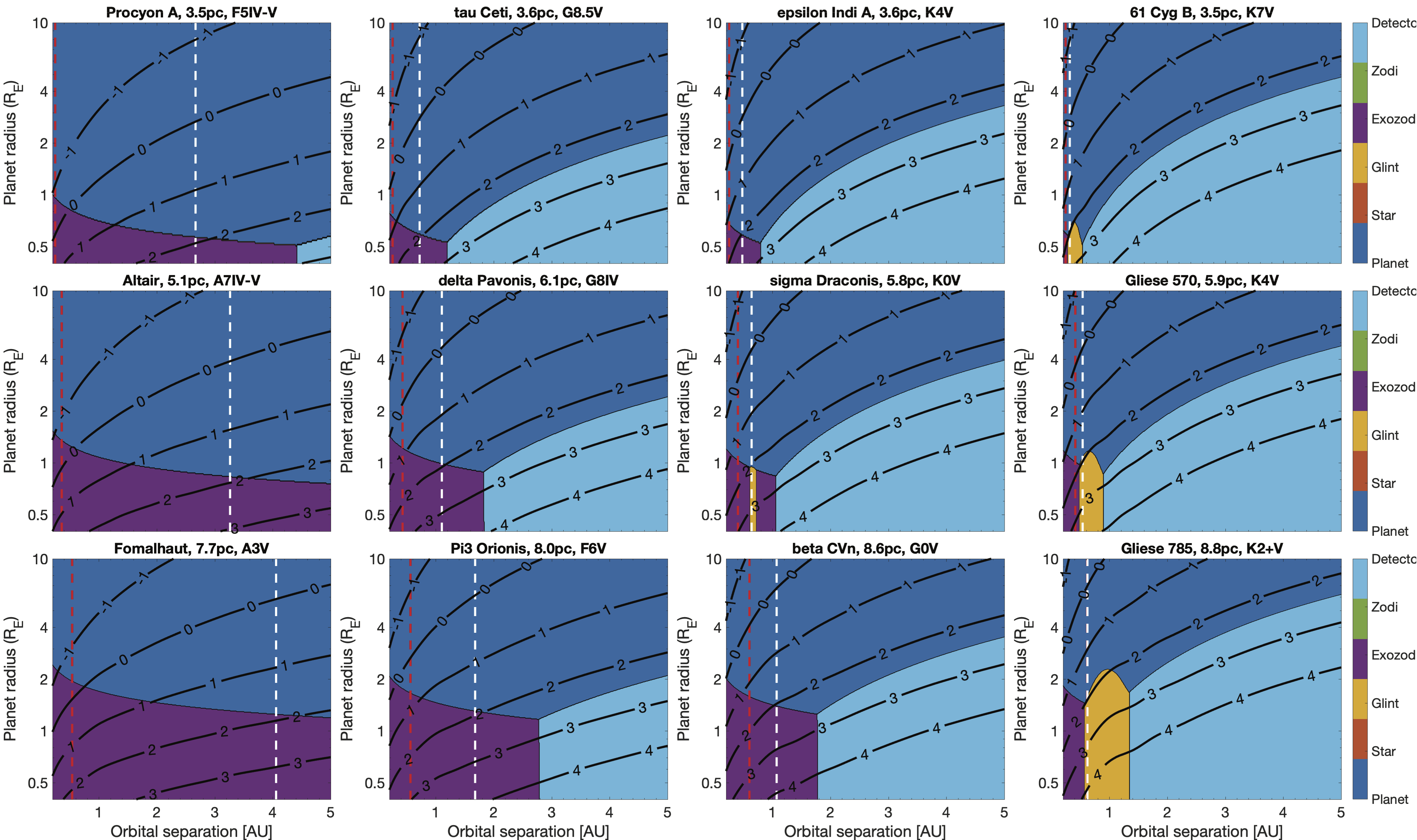}
\caption{Expected integration time (S/N=20 per spectral element, contour profiles, in log(hours)) and dominant noise source (shaded areas) expected in HabEx exploration of planets around nearby stars. The red dashed line corresponds to the IWA, and the blue dashed line corresponds to the planet-star separation for receiving Earth's insolation, i.e., the 1-AU equivalent. We adopt the nominal parameters in Table \ref{tab:parameter} in these simulations. With the additional optical edge coating\cite{McKeithen2020JATIS}, the impact of the solar glint would be eliminated.}
\label{fig:noise_h}
\end{figure}

Fig. \ref{fig:noise_h} shows the expected integration time and the underlying dominant noise source to detect planets around nearby stars with demonstrated starshade optical performance and telescope parameters that approximately correspond to the HabEx mission concept (Table \ref{tab:parameter}). Based on nominal performance parameters, solar glint would be the dominant noise term for observing Earth-sized and smaller planets in the habitable zones of the latest-type stars in each distance group. This is because the habitable zones of these stars approach the IWA, where the solar glint is the brightest. If the optical edge coating is applied and the solar glint brightness is reduced by a factor of 10, this impact of the solar glint would be eliminated.

We see that photon-limited detection and characterization of planets can be expected in the majority of the parameter space explored. Particularly, HabEx would be able to perform photon-limited planet characterization for Earth-sized planets in the habitable zones of nearby stars as far as $\sim6$ parsecs. The integration time needed to achieve an S/N=20 spectrum at the resolution of $R=140$ is $\sim10$ hours for the closest (3 -- 4 parsecs) stars and $50\sim100$ hours for stars 5 -- 6 parsecs away. These estimated integration times indicate more than one-order-of-magnitude improvement over the Starshade Rendezvous with {\it Roman}. The improvement comes from not only the larger telescope, but also higher throughput, better angular resolution to reduce the exozodiacal light interference, and lower solar glint brightness.

Exozodiacal light starts to affect the observations of Earth-sized planets around stars $\geq8$ parsecs away. Fig. \ref{fig:noise_h} shows that the exozodiacal light is the dominant noise term for planets $<1.5$ R$_\oplus$ in the habitable zones of $\sim8$-parsec stars. If the exozodiacal light can be subtracted to the photon-noise limit, however, the planets may still be observed and characterized with S/N=20 spectra within a few hundred hours. If the parameter $\alpha$ in Eq. \ref{eq:sn} is further reduced from two to unity, the integration time in the exozodiacal-dominant regime would be smaller than what is shown in Fig. \ref{fig:noise_h} by a factor $\sim2$.

The potential targets for HabEx include more distant stars up to $\sim15$ parsecs (Fig. \ref{fig:target}). While not explicitly shown in Fig. \ref{fig:noise_h}, we estimate the impact of solar glint and exozodiacal light on the performance of planet observations around these farther stars. For the photometric aperture adopted in this study (i.e., of the diameter $\lambda/D$), we estimate that the exozodiacal flux from a ``3 zodis'' disk has a magnitude of 28.3 at the habitable zone. The flux from the solar glint, with coating and 46\% encircled energy, has a magnitude of 30.4. The performance in each distance group will be characterized by the latest-type (i.e., the least luminous) stars in that group, and the habitable zones of these stars typically have the angular separation corresponding to the IWA (70 mas). In the case of Gliese 785, for example, the habitable zone is at $\sim0.62$ AU, and a 1-R$_{\oplus}$ planet would have a planet-star contrast of $8.5\times10^{-10}$. With the star's visual magnitude of 6 and the encircled energy of 46\% in the photometric aperture, the planet would have a magnitude 29.4. This is consistent with Fig. \ref{fig:noise_h}: a 1-R$_{\oplus}$ planet would be brighter than the solar glint but less bright than the exozodiacal light, and a 1.6-R$_{\oplus}$ planet would be approximately as bright as the exozodiacal light. Applying the same analysis to a 15-parsec star, the habitable zone is at $\sim1.05$ AU (derived from the IWA), and a 1-R$_{\oplus}$ planet would have a planet-star contrast of $2.6\times10^{-10}$. The star will have a similar apparent magnitude as Gliese 785 because its luminosity would scale as the square of the distance to keep the angular separation of the habitable zone at the IWA. The planet would then have a magnitude of 30.7 in the photometric aperture. This implies that for the most distant stars in HabEx's planet search, a 1-R$_{\oplus}$ planet would be as bright as the solar glint, and a 3-R$_{\oplus}$ planet would be as bright as the exozodiacal light. As such, HabEx with coating generally does not have solar glint as the main noise term and will be increasingly impacted by exozodiacal light for stars 8 -- 15 parsecs away.

\section{Discussion}
\label{sec:discussion}

\subsection{Requirements for Background Calibration}
\label{sec:calibration}

To this point we work on the assumption that the background can be removed by imaging processing to the photon-noise limit (i.e., $\beta=0$ in Eq. \ref{eq:sn}). Since the solar glint and the exozodiacal light dominate over the planetary light in the noise budget in substantial fractions of {\it Roman}'s and HabEx's search spaces for small planets (Figs. \ref{fig:noise}, \ref{fig:noise_g}, and \ref{fig:noise_h}), the ability to calibrate these noise terms is important for delivering the science capability described in Section \ref{sec:analysis}. 

\begin{figure}[!h]
\centering
\includegraphics[width=1.0\textwidth]{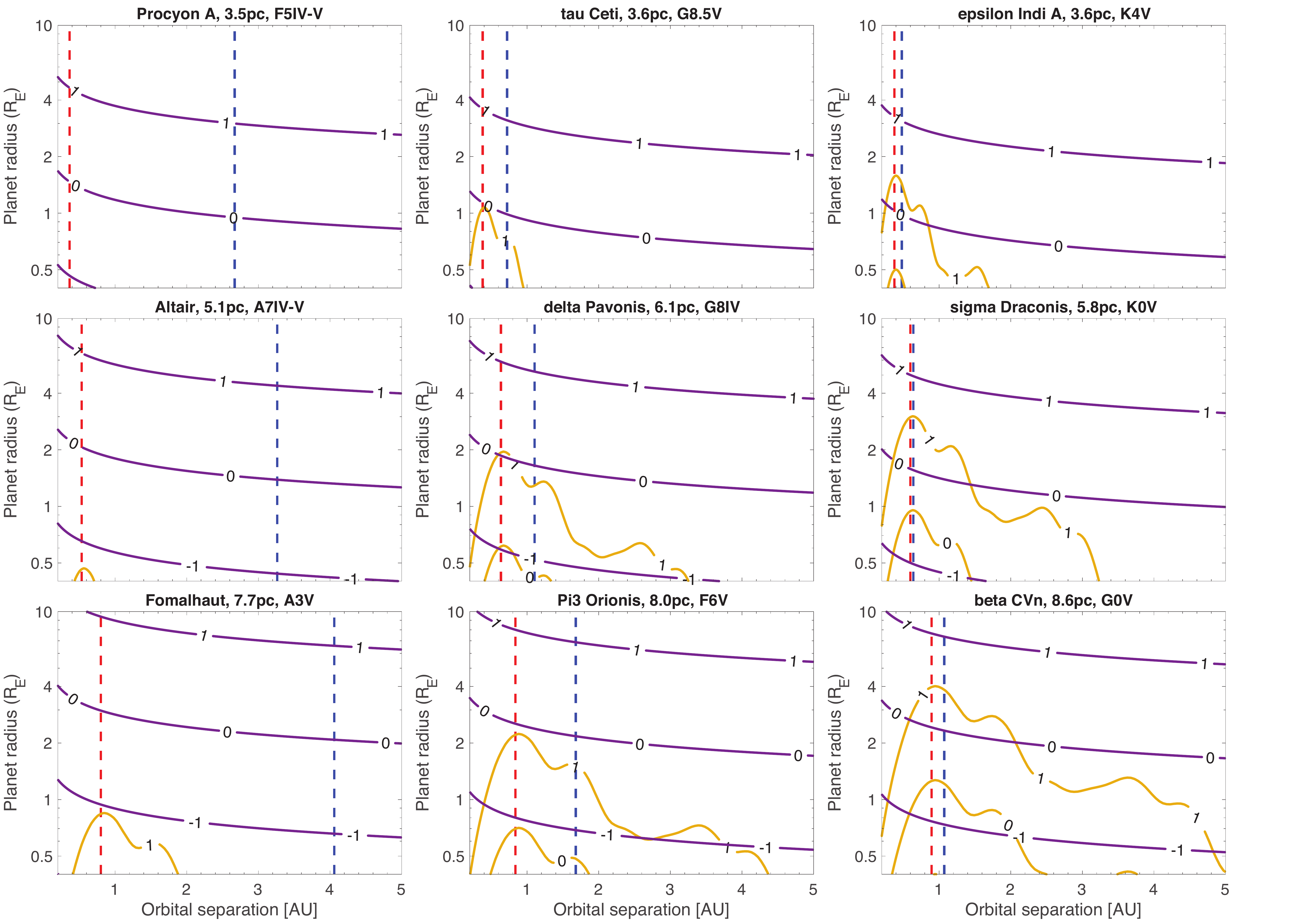}
\caption{The ratio between the planetary light and the solar glint (orange) and the ratio between the planetary light and the exozodiacal light (purple), for the same simulation parameters as Fig. \ref{fig:noise_g}. The base-10 logarithmic values are shown and labeled along the contour profiles.}
\label{fig:cali}
\end{figure}

Here we first quantify the precision needed for background calibration. Fig. \ref{fig:cali} shows the ratio between the planetary flux and the background flux from solar glint and exozodiacal light (i.e., $N_{\rm P}/\beta N_{\rm B}$, or $K$ in Eq. \ref{eq:sn1}). As discussed in Section~\ref{sec:background}, the maximum S/N achievable is $N_{\rm P}/\beta N_{\rm B}$ when $\beta$ is nonzero. Therefore, the ratio shown in Fig. \ref{fig:cali}, divided by the desired S/N (i.e., 20 in this work), is the maximum tolerable residual fraction ($\beta$) of the background after calibration.

For the closest stars ($<4$ parsecs), the flux from a 1-$R_\oplus$ planet in the habitable zone is similar to the flux of the exozodiacal light and is greater than the flux of solar glint by more than one order of magnitude (Fig. \ref{fig:cali}). This means that the exozodiacal light must be calibrated to 5\% to allow an S/N of 20. For farther stellar systems ($5\sim6$ parsecs), the requirement for the calibration precision of exozodiacal light becomes 2\%, and that for the calibration of solar glint is 5\%. For stars $\sim8$ parsecs away, the limiting science objective for a reasonable integration time with {\it Roman} would be to characterize 1.5-$R_\oplus$ planets in the habitable zones (Fig. \ref{fig:noise_g}). For this, the requirement for the calibration precision of exozodiacal light and solar glint remains 2\% and 5\%, respectively. Taken together, the residual exozodiacal light should be less than 2\% (i.e., $\beta_{\rm exozodi}\leq0.02$) and the residual solar glint should be less than 5\% (i.e., $\beta_{\rm glint}\leq0.05$) to avoid significant adverse impact on the spectral characterization of small and temperate planets. At these critical precision levels of background calibration, the final S/N would be degraded from the S/N with photon-noise-limit calibration by a factor of $\sqrt{2}$ (Eq. \ref{eq:sn1}), or $\sim14$. To achieve a final S/N of 20, the photon-noise-limit S/N should be $20\times\sqrt{2}\sim28$; this would be achievable by doubling the integration time from what is shown in Section~\ref{sec:analysis} and better precision for background calibration (i.e., $\sim1\%$ for exozodiacal light).

Would this level of calibration be achievable? For solar glint, which has a smooth intensity profile and only depends on the starshade position, orientation, and solar angle, standard image-processing techniques should be able to subtract it out with high precision. To enable this calibration, it may be necessary to carry out a reference observation of the solar glint after launch. A future community data challenge\cite{Hu2020JATIS} should develop and confirm the capability to subtract the solar glint with the help of the reference observation.

For exozodiacal light, the requirement for calibration is more stringent and the ability to subtract and remove it depends on the spatial smoothness of the dust disk. One resolution element of {\it Roman} at 700 nm is $\sim60$ mas, which corresponds to 0.2 -- 0.5 AU in the nearby planetary systems. Therefore, density fluctuation of exozodiacal disks at this scale may adversely impact the calibration and extraction of the embedded planetary signal. While there has not been a direct observation of the exozodiacal light in the habitable zones of nearby stars in the visible and near-infrared wavelengths, theoretical models of the origin of the dust particles may shed light on their distribution. In-situ production of the dust by collisions and giant impacts would lead to bright and localized dust ``clumps''\cite{kennedy2013bright}, but these events are rare in mature planetary systems, which is the case for most stars in the target lists of {\it Roman} and HabEx. Dust particles may also be transported to the habitable zones by Poynting-Roberson drag\cite{kennedy2015warm}, or delivered as comets scattered by outer planets and then sublimated\cite{bonsor2012scattering}. The latter two processes would eventually produce a smoothly varying dust density profile\cite{bonsor2018using}, and may enable high-precision background removal. However, dust particles may be trapped in mean-motion resonance with the planets\cite{stark2008detectability,stark2011transit} and become ``clumps.'' In addition to spatial structures, exozodiacal dust particles should have very different spectral shapes than the planets, which may provide another way to distinguish them. We stress that unlike other background terms, the exozodiacal signal and the density profile it implies is a science objective in its own right. Its spatial distribution and origin would be studied together with the planets.

\subsection{Variability of the Background}
\label{sec:variability}

Another potential challenging aspect of background calibration is that the background may vary during the long integration typically required for planet characterization. As demonstrate in Section~\ref{sec:analysis}, the solar glint is the dominant background term compared to other stray light sources and residual starlight. The solar glint brightness is expected to vary, as it depends on the lateral position of the starshade with respect to the line of sight. The next-in-the-line stray light source, reflection of earthshine and the Milky Way (Section~\ref{sec:brightbody}), can be variable as well\cite{jiang2018using}. The solar glint and other stray light to starshade exoplanet imaging is the analog of ``speckles'' to coronagraph exoplanet imaging, and the variability of speckles has driven the design of the coronagraph instrument on {\it Roman}\cite{nemati2017sensitivity}. Here we provide a high-level estimate on the impact of background variability on starshade exoplanet imaging.

\begin{figure}[!h]
\centering
\includegraphics[width=1.0\textwidth]{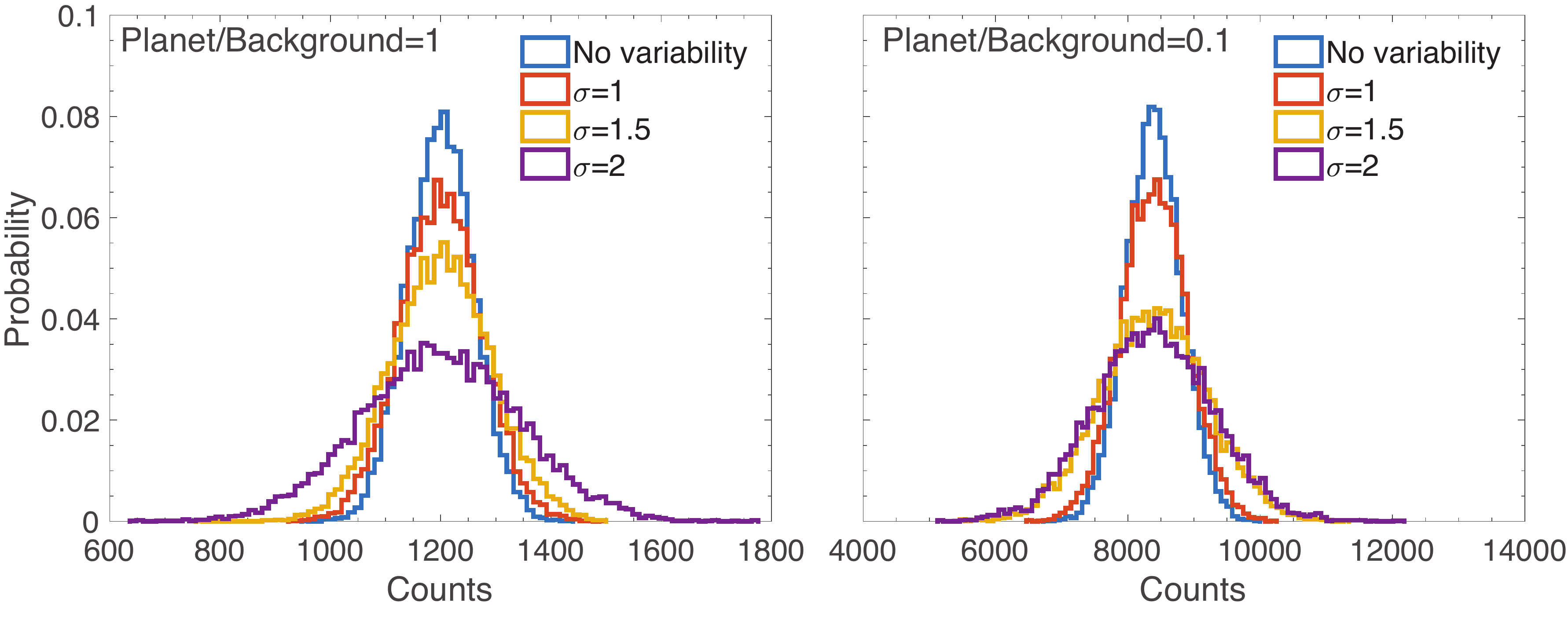}
\caption{Effect of background variability on image subtraction and planet detection. For each scenario, 10,000 instances of observations are simulated. Each observation consists of photometry of two apertures, one with the background and the other with the background and the planet, followed by subtraction of one aperture from the other. Each observation is divided into 100 segments, and the background flux in each segment varies around the median flux and follows a log-normal distribution with $\sigma$ shown in the figure. The counts of the apertures in each segment follow a Poisson distribution with a mean of the background flux of that segment (and the planet flux if any). The standard deviation of the resulting distribution characterizes the uncertainty of the planet flux measurement. In both cases, the median counts are chosen so that S/N=20 of the planet measurement is achieved at the limit of no variability.}
\label{fig:var}
\end{figure}

We approximate the background variability as a random variation, because the variation of the starshade's lateral position would have a shorter timescale ($<10$ minutes\cite{Flinois2018}) compared to typical integration times. A change in the starshade's position is equivalent to a change in the off-axis angle for this problem, and the brightness of the residual starlight and the solar glint would have a \textit{fractional} change as a function of the off-axis angle (Fig. \ref{fig:scaling}). We thus approximate the background flux with a log-normal distribution with the parameter 0 and $\sigma$ (${\rm Lognormal}(0,\sigma)$) with respect to the median value. We perform Monte Carlo simulations of photons arriving onto two photometric apertures on the detector. Both apertures receive photons from the varying background and one of the pixel additionally receive photons from the planet. The count difference between the two apertures -- and the statistical distribution of it -- thus indicates how well the planet is detected. Fig. \ref{fig:var} shows the resulting distributions for different levels of variability.

First of all, we see that the mean count in the subtracted aperture remains the same with and without the variability in Fig. \ref{fig:var}. This indicates that the random variability would not bias the measurement of the planetary flux.

Fig. \ref{fig:var} also shows that the resulting distribution becomes wider with greater background variability, indicating that background variability degrades the S/N of the planet measurement. When the planetary flux is comparable to the background flux, the standard deviation of the resulting distribution increases from that of an invariant scenario by 11\%, 62\%, and 106\% when the value of $\sigma$ is 1, 1.5, and 2, respectively. If the planetary flux is one order of magnitude less than the background flux, the standard deviation increases by 22\%, 91\%, and 167\%, respectively. Therefore, a random variability that follows a log-normal distribution up to $\sigma=1$ would not cause an appreciable change in the planet S/N, but a more variant background would significantly degrade the S/N, especially when the image is background-dominated. 

To put this into perspective, a log-normal distribution of $\sigma=1$ would imply that the background flux stays within approximately half an order of magnitude from the median value in 68\% time during observation. This allowed range of variability would correspond to a variation of 20 mas in the off-axis angle (Fig. \ref{fig:scaling}) or 2.5-m in the lateral displacement in the case of Starshade Rendezvous with {\it Roman}. The demonstrated capability of formation flying should be able to control the starshade to stay well within this range during science operation\cite{Flinois2018}. The allowed range of variability (i.e., half an order of magnitude) appears to be also greater than the variability of earthshine\cite{jiang2018using}. Therefore, the variation in the solar glint and residual starlight caused by the starshade's motion in formation flying, as well as the variation in the brightness of the earthshine, should not result in substantial degradation of the S/N of planet measurements.

% \subsection{Sensitivity to the Size of Telescope}

% Omitting detector systematics, Fig. \ref{fig:snr_zodi} shows a reasonable scenario of directly imaging and characterizing planets of a star 10 parsecs away. In order to detect the spectral feature of O$_2$ in the atmosphere, for a reasonable integration time of ~20 days, an Earth-sized planet requires a 5.2-m telescope, while a 2-R$_\oplus$ planet only requires a 2.6-m telescope. In addition, if we require that the main astrophysical background, $N_{\rm E}$, would be no more than a factor of 10 greater than the planetary signal NP for detection, an Earth-sized planet requires a 3.0-m telescope, while 2-R$_\oplus$ planet only requires a 1.4-m telescope. The simple estimates provided here point out a basic fact that directly imaging Earths and super-Earths are limited by exozodiacal dust.

% \begin{figure}[!h]
% \centering
% \includegraphics[width=0.5\textwidth]{figure/snr_zodi_diameter.eps}
% \caption{(Top) Required integration time to detect (S/N=5) and characterize (S/N=25) an Earth-sized and 2-Earth-sized exoplanet of at 1-AU orbit of a Sun-like star 10 parsecs away, at 700 nm for a wavelength bin of 10 nm. (Bottom) The ratio between the exozodiacal light and the planetary light for the same parameters. The planetary albedo including phase function is 0.2, exozodi dust density is 5 times with respect to the Solar System\cite{Stark2014,Ertel2020}, the contrast is $10^{-10}$, and the end-to-end efficiency of the optical system is 0.05.}
% \label{fig:snr_zodi}
% \end{figure}

\subsection{Sensitivity of Exozodi Levels}
\label{sec:discusszodi}

We have assumed the exozodi level to be 3 zodis in this work, guided by the best-fit median value from the hitherto most sensitive exozodiacal dust survey\cite{Ertel2020}. However, the survey showed a large spread in the amount of exozodi light in nearby stars, with the 95\% upper limit of 27 zodis\cite{Ertel2020}. With a larger exozodi level, observations in a greater search space of planetary size and orbital separation (Figs. \ref{fig:noise}, \ref{fig:noise_g}, and \ref{fig:noise_h}) will be dominated by exozodiacal light. In the exozodi-dominated regime, the required integration time and the precision for background calibration will increase linearly as the exozodi level increases (Eqs. \ref{eq:sn} and \ref{eq:sn1}).

Of particular interest is where nearly photon-limited planetary spectroscopy may be expected with good background calibration. If the exozodi level increases from 3 zodis to 10 zodis, the boundary between the planet-dominated regime and the exozodi-dominated regime would increase from $\sim1\ R_{\oplus}$ to $\sim1.8\ R_{\oplus}$ for \textit{Roman} observing the nearest ($<4$ parsecs) stars (Fig. \ref{fig:noise_g}). The boundary would increase to $>3\ R_{\oplus}$ for farther stars. This means that an increase of the exozodi level could prevent \textit{Roman} Rendezvous to perform photon-limited planetary spectroscopy for any target stars. HabEx on the other hand is less susceptible to high exozodi levels. If the exozodi level increases from 3 zodis to 10 zodis, HabEx would still be able to obtain photon-limited planetary spectroscopy for Earth-sized or larger rocky planets ($<1.7\ R_{\oplus}$) around stars $<6$ parsecs away (Fig. \ref{fig:noise_h}). The density and structure of the exozodiacal disks of nearby stars will continue to be an important area of exploration.

\subsection{Implications on Mission Designs}

The analyses presented in this paper assume that the observations of the planets would take place at the orbital phase angle of $\pi/3$. In reality, we do not know planets and their orbital elements of most target stars. The design of future missions using starshades must consider the ``search completeness'' that factors in the randomness of the observation epoch in planets' orbital revolution and visibility. Both Starshade Rendezvous with {\it Roman} and HabEx include an essentially blind search of planets around the nearby stars\cite{Seager2019,Gaudi2020}, and the search completeness of Earth-sized planets in the habitable zone was the driving factor of design and science cases.

However, the feasibility of spectral characterization shown in this paper paints a remarkably consistent picture with the search completeness, even though they are very different metrics. The search completeness primarily concerns about the ability to detect planets in the broadband with several temporally spaced visits, while spectral characterization may entail a long integration performed at a single visit. Fig. \ref{fig:noise_g} shows that characterizing Earth-sized and temperate planets would be feasible around the nearest (3 -- 4 parsecs) stars, and only marginally feasible around slightly farther (5 -- 6 parsecs) stars, for Starshade Rendezvous with {\it Roman}. This is fully consistent with the search completeness estimate, where the overall habitable-zone characterization completeness would be $>25\%$ for the nearest stars, and $\sim10\%$ or less for the slightly farther stars\cite{RomeroWolf2020JATIS}. In other words, a more favorable target for spectral characterization shown in this paper generally has more favorable search completeness.

Interestingly, the nearest stars (Procyon~A, $\tau$ Ceti, $\epsilon$~Indi~A, and Sirius~A, exhaustively) are truly the outstanding targets for both Starshade Rendezvous with {\it Roman} and HabEx, for the high search completeness, the relative loose requirement of background calibration, and short integration time to characterize Earth-sized planets in their habitable zones. With a starshade, {\it Roman} could already measure spectra and characterize the atmospheres of Earth-sized planets of these stars. HabEx would further provide the wide spectral coverage to pinpoint atmospheric abundance, and with an integration of hours for each spectrum, the possibility to measure the variability of the spectra. The spectral variability would inform surface compositions (e.g., land and sea) and indicate variable cloud coverage and hydrological cycles\cite{cowan2009alien,fujii2013variability,jiang2018using} --- this feat would be feasible only for the nearest stars using HabEx. Among the four stars, however, an outer dust disk has been detected with far-infrared and radio observations around $\tau$~Ceti\cite{greaves2004debris,lawler2014debris,macgregor2016alma}, and Procyon~A and Sirius~A have white dwarf companions whose evolution might have adversely impacted habitability of any planets nearby. This leaves $\epsilon$~Indi~A apparently be the most promising target. Because of this revelation, we suggest that mission designs should set a high priority to maximize the search completeness of planets in the habitable zones of these four stars, and particularly $\epsilon$~Indi~A, and encourage precursor efforts such as radial-velocity measurements that prioritize these stars.

We have used the reference wavelength of 700 nm for the analyses presented in this work, and it provides a representation of the ``green'' band of Starshade Rendezvous with {\it Roman} (615-800 nm\cite{RomeroWolf2020JATIS}) and the UV-Visible band of HabEx (0.3--1.0 $\mu$m\cite{Gaudi2020}). \textit{Roman} may also have imaging and spectroscopy capabilities in the ``blue'' band (425-552 nm\cite{Seager2019}). The blue band offers a few advantages over the green band, with less bright solar glint by 2 magnitudes\cite{Hilgemann2019} and a smaller PSF and thus less exozodiacal light by a factor of $\sim2$. These advantages would reduce the needed integration time to achieve S/N and loosen the requirements on background calibration. The blue band however requires a larger separation between the starshade and the telescope (37.2 Mm compared to 25.7 Mm for the green band), leading to longer distances and higher fuel and time cost to maneuver the starshade from target to target. Also, the blue band likely has fewer spectral features of interest from exoplanet atmospheres than the green band\cite{Lupu2016,Nayak2017,feng2018,Hu2019B2019ApJ...887..166H,damiano2020exorel,damiano2020multi}. In addition, HabEx would perform near-infrared spectroscopy (1.0--1.8 $\mu$m) on selected ``high-interest'' targets\cite{Gaudi2020}. The magnitude of the solar glint has not been evaluated for the HabEx starshade in the near-infrared band, nor has the optical edge coating been designed to cover the near-infrared band. Future studies are required to quantify the solar glint and its impact on the science performance of HabEx in the 1.0--1.8 $\mu$m band.

Lastly, the planet search space that would be fully accessible for Starshade Rendezvous with {\it Roman} would be large rocky planets ($\sim1.5$ $R_{\oplus}$) and potentially water worlds ($\sim2.5$ $R_{\oplus}$). For all stars in the target lists, these planets would likely be within the reach from the perspectives of integration time and background calibration. The search completeness has not been assessed for these ``super-Earths'', but we suspect that they would have reasonable completeness. Therefore, detailed designs of Starshade Rendezvous with {\it Roman} would need to prioritize searching and characterizing large rocky planets and water worlds in the habitable zones, and relevant science investigations would need to further quantify the requirements for their orbital determination and atmospheric abundance retrieval.

\section{CONCLUSION AND PROSPECTS}
\label{sec:conclusion}

We provide an overview and reassessment of the noise budget of exoplanet imaging and spectroscopy enabled by a starshade in formation flight with a space telescope. We have developed a framework to estimate the S/N of the planet observations of nearby stars, using demonstrated performance parameters of starlight and stray light suppression resulted from S5 work. With an analysis of miscellaneous sources of stray light -- from leakage through micrometeoroid damage to the reflection of earthshine -- we show that the dominant stray light source is the scatter of sunlight by the edge of the starshade, i.e., the solar glint. The solar glint is a starshade-unique noise source and the starshade's analog to coronagraph's speckles.

Applying our analysis framework to Starshade Rendezvous with {\it Roman} and HabEx, we find that starlight suppression delivered by the starshades will be well enough to eliminate residual starlight from the dominant noise term, and the optical edge coating technology shown in Ref. \cite{McKeithen2020JATIS} would be necessary to prevent the solar glint from becoming the dominant noise. For a uniform dust level of 3 zodis, the dominant noise term is likely exozodiacal light for characterizing Earth-sized planets around stars $>4$ parsecs away with {\it Roman} and $>7$ parsecs away with HabEx. For closer stars, these missions with starshades would provide photon-limited measurements of Earth-sized planets, if unbiased calibration of the background to the photon-noise limit can be achieved with reference observations and image processing.

Considering holistically the number of accessible stars, integration times, and demands for precise background calibration, one may expect Starshade Rendezvous with {\it Roman} to probe the nature of temperate and large rocky planets ($\sim1.5$ R$_{\oplus}$) and HabEx to study the nature of Earth-sized planets and find true Earth twins. Based on the S/N estimates presented in this paper, {\it Roman} with a starshade would be capable of obtaining high-precision, moderate-resolution spectra of temperate and large rocky planets for stars as far as $\sim8$ parsecs with a reasonable observation time, while spectroscopy of Earth-sized planets would likely be limited to the nearest few stars. HabEx, with not only a larger telescope but also better angular resolution (to reduce exozodiacal light) and a more distant starshade (to reduce solar glint), would drastically reduce the required observation time, and make possible to characterize Earth-sized planets for stars $\sim8$ parsecs and even farther away.

Based on the analyses presented here, we find it essential to validate the optical edge coating technology to eliminate the adverse impact of the solar glint on the science performance of Starshade Rendezvous with {\it Roman} and HabEx. Also, as most scenarios of planet detection and characterization require precise calibration of the background, mostly from exozodiacal light, it would be essential to develop the imaging processing techniques for the background calibration and validate their ability using realistic simulated images, for example, in a data challenge\cite{Hu2020JATIS}. In all, with unprecedented knowledge of starshade's optical performance and maturity of the associated technologies, we confirm that a starshade coupled with a sizeable space telescope continues to provide a near-term pathway towards finding habitable Earths in our interstellar neighborhood.

%%%%%%%%%%%%%%%%%%%%%%%%%%%%%%%%%%%%%%%%%%%%%%%%%%%%%%%%%%%%%
\acknowledgments     %>>>> equivalent to \section*{ACKNOWLEDGMENTS}
We thank Sergi Hildebrandt for providing example SISTER simulations, and Karl Stapelfeldt, Eric Mamajek, and Andrew Romero-Wolf for helpful discussion. The research was carried out at the Jet Propulsion Laboratory, California Institute of Technology, under a contract with the National Aeronautics and Space Administration (80NM0018D0004).

%%%%% References %%%%%

\bibliography{report}   % bibliography data in report.bib
\bibliographystyle{spiejour}   % makes bibtex use spiejour.bst

\section*{Author Biography}

{\bf Renyu Hu} is a scientist at the Jet Propulsion Laboratory and his research strives to identify and characterize habitable environments in the Solar System and beyond. He is the Starshade Scientist of the NASA Exoplanet Exploration Program, providing science leadership to the S5 Starshade Technology Development Activity and managing the Starshade Science and Industry Partnership program. He received his PhD in Planetary Science at Massachusetts Institute of Technology in 2013.

\noindent{\bf Doug Lisman} is the systems engineering lead for starshade technology development at the Jet Propulsion Laboratory where he is a member of the Instrument Systems Engineering group. He received his BS degree in mechanical engineering from Washington University in St. Louis in 1984 and has been at JPL since 1984.

\noindent{\bf Stuart Shaklan} is the supervisor of the High Contrast Imaging Group in the Optics Section of the Jet Propulsion Laboratory. He received his Ph.D. in Optics at the University of Arizona in 1989 and has been with JPL since 1991.

\noindent{\bf Stefan Martin} is a senior optical engineer at the Jet Propulsion Laboratory. He received his BSc degree in physics from the University of Bristol, United Kingdom, and his PhD in engineering from the University of Wales. At JPL, he has been leader of the TPF-I Flight Instrument Engineering Team, testbed lead for the TPF-I Planet Detection Testbed, and payload lead for the HabEx Telescope design study. He is currently involved in starshade accommodation on future space telescopes, such as NGRST.

\noindent\textbf{Phil Willems} is an Optical Engineer at the Jet Propulsion Laboratory, where he is the manager of the S5 Starshade Technology Development Activity. He received his BS degree in physics from the University of Wisconsin- Madison in 1988, and his PhD degree in physics from the California Institute of Technology in 1997.

\noindent\textbf{Kendra Short} is currently the Deputy Flight System Manager for NASA’s Europa Clipper mission. Prior to this, she was the Deputy Program Manager for the NASA Exoplanet Exploration Program, including S5 Starshade Technology Development Activity. Ms. Short received a B.S. in Mechanical \& Aerospace Engineering from Princeton University and an M.S. in Aero/Astro Engineering from Stanford University.

\listoffigures
\listoftables

\end{spacing}
\end{document}